\documentclass{article}
\usepackage{jheppub}

\pdfoutput=1
\usepackage{amsmath,amssymb,graphicx}
\usepackage{longtable}
\usepackage{verbatim}
\usepackage{float}

\usepackage[usenames,dvipsnames]{xcolor}
\definecolor{darkerblue}{rgb}{0.2,0.2,0.5}

\usepackage{tikz}
\usetikzlibrary{trees}
\usetikzlibrary{decorations.pathmorphing}
\usetikzlibrary{decorations.markings}
\tikzset{
    photon/.style={decorate, decoration={snake}, draw=black},
    wino/.style={draw=redwine},    
    electron/.style={draw=black, postaction={decorate},
        decoration={markings,mark=at position .55 with {\arrow[draw=black]{>}}}},
    scalar/.style={draw=black, dashed,postaction={decorate},
        decoration={markings,mark=at position .55 with {\arrow[draw=black]{>}}}},
    gluon/.style={decorate, draw=black,
        decoration={coil,amplitude=4pt, segment length=5pt}}
}

\usepackage{afterpage}

\newcommand{\mst[1]}{m_{{\tilde t}_#1}}
\newcommand{\msb[1]}{m_{{\tilde b}_#1}}
\newcommand{\thetast}{\theta_{\tilde t}}
\newcommand{\thetasb}{\theta_{\tilde b}}

\newcommand{\bear}{\begin{array}}
\newcommand{\ear}{\end{array}}

\newcommand{\beq}{\begin{eqnarray}}
\newcommand{\eeq}{\end{eqnarray}}
\newcommand{\beqa}{\begin{eqnarray}}
\newcommand{\eeqa}{\end{eqnarray}}

\def\OMIT#1{{}}
\newcommand{\lsim}{\mathrel{\rlap{\lower4pt\hbox{\hskip1pt$\sim$}}
    \raise1pt\hbox{$<$}}}         %less than or approx. symbol
\newcommand{\gsim}{\mathrel{\rlap{\lower4pt\hbox{\hskip1pt$\sim$}}
    \raise1pt\hbox{$>$}}}         %greater than or approx. symbol

\begin{document}

\title{\bf \color{darkerblue} Precision Natural SUSY at CEPC, FCC-ee, and ILC}

\author[a]{JiJi Fan,} 
\author[b]{Matthew Reece}
\author[c]{and Lian-Tao Wang}
\affiliation[a]{Department of Physics, Syracuse University, Syracuse, NY, 13210, USA}
\affiliation[b]{Department of Physics, Harvard University, Cambridge, MA 02138, USA}
\affiliation[c]{Enrico Fermi Institute and Kavli Institute for Cosmological Physics, \\University of Chicago, Chicago, IL 60637, USA}

\abstract{Testing the idea of naturalness is and will continue to be one of the most important goals of high energy physics experiments. It will play a central role in the physics program of future colliders. In this paper, we present projections of the reach of natural SUSY at future lepton colliders: CEPC, FCC-ee and ILC. We focus on the observables which give the strongest reach, the electroweak precision observables (for left-handed stops), and Higgs to gluon and photon decay rates (for both left- and right-handed stops). There is a ``blind spot'' when the stop mixing parameter $X_t$ is approximately equal to the average stop mass. We argue that in natural scenarios, bounds on the heavy Higgs bosons from tree-level mixing effects that modify the $h b {\bar b}$ coupling together with bounds from $b \to s\gamma$ play a complementary role in probing the blind spot region.  For specific natural SUSY scenarios such as folded SUSY in which the top partners do not carry Standard Model color charges, electroweak precision observables could be the most sensitive probe. In all the scenarios discussed in this paper, the combined set of precision measurements will probe down to a few percent in fine-tuning.}
\maketitle

%%%%%%%%%%%%%%%%%%%%%%%%%%%%%%%%%%
\section{Introduction}
\label{sec:intro}
%%%%%%%%%%%%%%%%%%%%%%%%%%%%%%%%%%

Naturalness is an interesting theoretical idea which has motivated our expectation for new physics beyond the Standard Model. At the same time, as a potential candidate framework for new physics, supersymmetry has many appealing features. The convergence of these two ideas has made natural SUSY one of the prime targets of the new physics searches at the LHC.  The most robust prediction of natural SUSY is the presence of light stops~\cite{Dimopoulos:1995mi,Cohen:1996vb,Kitano:2006gv,Perelstein:2007nx}. They have not been discovered yet at the LHC Run 1. Such a pursuit will continue to be a central theme of the physics program of Run 2. Due to its importance, we will not give up even if HL-LHC turns up empty. Therefore, it will be a top target for the future colliders as well.\footnote{In this paper, we will not distinguish between the {\it naturalness} and the level of (absence of) {\it fine-tuning} in a theory, and we will use these two terms interchangeably.}

The null result for the stop searches could be due to two types of reasons. The stop could be heavier than the kinematical reach of the collider. In this case, the reach of future lepton colliders is limited by the relatively low center of mass energy. Only CLIC can in principle have a reach comparable to that of the HL-LHC. The high energy proton colliders, such as the hadronic mode of the Future Circular Collider  (FCC-hh) and the Super proton-proton Collider (SppC), have better prospects. On the other hand, stops can also be hidden due to some ``non-standard" decay modes and/or kinematics of the decay products~\cite{Barbier:2004ez,Csaki:2011ge,Strassler:2006im,Strassler:2006qa,Han:2007ae,Fan:2011yu,Fan:2012jf,LeCompte:2011cn,LeCompte:2011fh,Graham:2012th,Graham:2014vya,Alves:2013wra}. In this case, precision measurements at lepton colliders could provide complementary probes independent of the details of stop decays. In this paper, we focus on the potential of future lepton colliders in covering the gaps of searches at high energy proton colliders and raising the ``bottom line" of the test of naturalness. 

There are two classes of precision observables at future $e^+e^-$ colliders which are sensitive to the presence of light stops. First, due to their significant coupling to the Higgs boson, the stops can induce considerable shifts in some Higgs couplings at the one-loop level. Among them, we observe that the constraints from $h \gamma \gamma$ and $h g g$ measurements are stronger than that from the Higgs boson wave-function renormalization in most cases. The other class is the electroweak precision test (EWPT) observables. Among them, the $T$-parameter is the most sensitive observable to the left-handed stops. A main conclusion of this paper is that future lepton colliders can push the test of naturalness to the level of a few percent by a combination of these observables. 

There is also an interesting ``blind spot" in the stop parameter space in which the coupling between the lightest stop and the light Higgs boson vanishes. In this region, the observables discussed above are much less sensitive. We note that in this case, the level of fine-tuning is controlled by the mass of the heavier stop. Hence, hiding a light stop does not lead to a much more natural theory. In addition, the light stop is not completely hidden from the full suite of precision measurements. In particular, the measurement of $b \to s \gamma$ could be useful. The stop contribution to this process also depends on $\tan\beta$ and $\mu$, parameters in the Higgs sector. The implication of this measurement for naturalness also depends on the value of $m_A$~\cite{Katz:2014mba}. Tweaking these parameters can loosen the limits on the stop. Yet some of these parameters are also constrained by the $h b \bar{b}$ coupling measurement. A combination of these two probes can push the level of the test of naturalness in the blind spot region to the several percent level as well.

The study of the physics potential of future $e^+ e^-$ colliders has a long history~\cite{Baur:1996zi,Gunion:1996cn, Heinemeyer:1999zd, Hawkings:1999ac}. An early study of the GigaZ prospects for constraining stops, albeit only for restricted subsets of the MSSM parameter space, appeared in Ref.~\cite{Erler:2000jg}. We will mostly rely on more recent studies for the ILC and FCC-ee~\cite{Baer:2013cma,Gomez-Ceballos:2013zzn,Asner:2013psa,Dawson:2013bba,Baak:2013fwa}.

Due to the null result in the stop searches, in recent years, alternative proposals in which top partners carry different gauge quantum numbers than the top have attracted renewed attention. To this end, we have generalized our study to the ``folded SUSY" scenario~\cite{Burdman:2006tz}. Folded stops have Standard Model electroweak quantum numbers but no charge under SU(3)$_c$. We observe that the $T$-parameter measurement could be the most sensitive probe of this scenario, except for the blind spot region, which is difficult to probe with any indirect observable.

Among the proposed lepton colliders FCC-ee, with its higher projected integrated luminosity, provides the best limit in Higgs coupling measurements. At the same time, the limits from CEPC are similar to those obtained from the ILC 1TeV scenario. With potential improvements suggested in~\cite{Fan}, the CEPC could be comparable to FCC-ee in the reach of EWPT.

In Section~\ref{sec:loops}, we discuss the parametric size of the leading loop effects to the electroweak and Higgs observables in natural SUSY and demonstrate how they arise as effective operators when stops and higgsinos are integrated out. In Section~\ref{sec:oblique}, we compute and present oblique EWPT constraints from future electron colliders on the stop sector. In Section~\ref{sec:Rb}, we compute and present constraints on the stop sector from the non-oblique observable $R_b$ at future electron colliders. In Section~\ref{sec:higgscoupling}, we compute and present Higgs coupling constraints on the stop sector from future electron colliders. In Section~\ref{sec:blindspot}, we discuss the physical origin of the blind spot region in the stop parameter space at future lepton colliders. In Section~\ref{sec:discussions}, we discuss probing the blind spot region by a combination of $b \to s \gamma$ and $hb\bar{b}$ coupling measurements. Then we present the key projections of this paper in Figure~\ref{fig:summaryplot} to show that the combined set of precision measurements will probe down to a few percent in fine-tuning. In the end, we comment that EWPT could be the most sensitive probe in a large bulk of the folded stop parameter space. 

%%%%%%%%%%%%%%%%%%%%%%%%%%%%%%%%%%
\section{Loop Effects of Natural SUSY}
\label{sec:loops}
%%%%%%%%%%%%%%%%%%%%%%%%%%%%%%%%%%

We would like to understand how $e^+ e^-$ colliders can constrain natural supersymmetric scenarios. Requiring a low degree of fine tuning imposes upper bounds on the masses of higgsinos, stops, and gluinos due to their respective tree-level, one-loop, and two-loop effects on electroweak symmetry breaking~\cite{Barbieri:1987fn,Dimopoulos:1995mi,Pomarol:1995xc,Cohen:1996vb,Kitano:2006gv,Perelstein:2007nx}. Because gluinos carry only SU(3)$_c$ quantum numbers, their effect on lepton collider processes is generally at a higher loop order than the effect of stops or higgsinos, which carry electroweak quantum numbers. Thus, we focus on understanding the dominant corrections to the Standard Model effective Lagrangian from integrating out stops and higgsinos. We assume that $R$-parity violation is small, in which case the leading corrections are always at one loop rather than tree-level. Furthermore, the largest contributions are generally those where the coupling appearing in the loop diagrams is the top Yukawa coupling $y_t \approx 1$. These include the $F$-term potential terms $\left|y_t H_u \cdot {\tilde Q}_3\right|^2 + \left|y_t H_u {\tilde u}^c_3\right|^2$.

In this section we will discuss the parametric size of the leading loop effects and demonstrate how they arise as effective operators when stops and higgsinos are integrated out. The discussions here help to understand the qualitative features of the results presented in later sections. In obtaining our numerical results, we will include the full loop functions as computed in the older literature, which are valid for arbitrary masses and mixings.

\subsection{Parametrization of Natural SUSY}

The stop mass-squared matrix, in the gauge eigenstate basis $(\tilde{t}_L, \tilde{t}_R)$, is given by 
\[ {\cal{M}}_{\tilde{t}}^2= \left( \begin{array}{cc}
m_{{\tilde Q}_3}^2+m_t^2 + \Delta_{\tilde{u}_L} & m_t X_t  \\
m_t X_t^* & m_{{\tilde u}_3}^2+m_t^2+\Delta_{\tilde{u}_R} \end{array} \right),\]
where $m_{{\tilde Q}_3}^2, m_{{\tilde u}_3}^2$ are the soft mass squared of left- and right- handed stops respectively and the stop mixing term $X_t = A_t - \mu/\tan\beta$. For simplicity, we will neglect possible phases in the stop mass matrix. The $D$-term quartic interactions give terms $\Delta_{\tilde{u}_R}=\left(\frac{2}{3} \sin^2\theta_W\right)\cos(2\beta)m_Z^2$ and $\Delta_{\tilde{u}_L}=\left(\frac{1}{2} - \frac{2}{3}\sin^2\theta_W\right) \cos(2\beta) m_Z^2$ which are $\ll m_t^2$. 

The stop mixing angle is related to the physical stop masses and mixing as 
\beq \label{eq:mixingangle}
\sin (2\theta_{\tilde{t}})= \frac{2 m_t X_t}{m_{\tilde{t}_2}^2-m_{\tilde{t}_1}^2}.
\eeq
We will choose $\theta_{\tilde t} \subset (-\pi/4, \pi/4)$ so the mass eigenstate with eigenvalue $m_{\tilde{t}_1}$ is mostly left-handed and the other one with $m_{\tilde{t}_2}$ is mostly right-handed. Not all possible values of $m_{\tilde{t}_2}^2$, $m_{\tilde{t}_1}^2$, and $X_t$ are allowed. In particular, for non-zero $X_t$, the region around $|m_{\tilde{t}_1}^2 - m_{\tilde{t}_2}^2| \sim 0$ may not be obtainable from the diagonalization of a Hermitian stop mass matrix~\cite{Fan:2014txa}.

The sbottom sector has a similar mass matrix with $m_t$ replaced by $m_b$, $m_{{\tilde d}_3}$ replacing $m_{{\tilde u}_3}$, and the appropriately modified $D$-terms. Generally we can neglect mixing in the sbottom sector because $m_b \ll m_t$. The mass of the left-handed sbottom $m_{\tilde{b}_1}^2$ could be written in terms of the stop physical masses and mixing angle as
\beq
m_{\tilde{b}_1}^2 = \cos^2\theta_{\tilde{t}} m_{\tilde{t}_1}^2 + \sin^2\theta_{\tilde{t}} m_{\tilde{t}_2}^2-m_t^2-m_W^2\cos(2\beta).
\eeq

In the higgsino sector, there are two neutral Majorana fermions and one charged Dirac fermion, with masses approximately equal to $\mu$. The splittings originate from dimension five operators when the bino and wino are integrated out, and are of order $m_Z^2/M_{1,2}$. We will ignore these splittings and treat all higgsino masses as equal to $\mu$ for the purpose of calculating loop effects.

\subsection{Electroweak Precision: Oblique Corrections}
\label{sec:loopsST}

The familiar $S$ and $T$ oblique parameters~\cite{Peskin:1991sw, Peskin:1990zt} (see also~\cite{Kennedy:1988sn, Holdom:1990tc, Golden:1990ig}) correspond, in an effective operator language (reviewed in ref.~\cite{Han:2004az,Han:2008es}), to adding to the Lagrangian
\beq
{\cal L}_{\rm oblique} = S \left(\frac{\alpha}{4 \sin \theta_W \cos \theta_W v^2}\right) h^\dagger W^{i \mu \nu} \sigma^i h B_{\mu \nu} - T \left(\frac{2 \alpha}{v^2}\right) \left|h^\dagger D_\mu h\right|^2. \label{eq:Loblique}
\eeq
Here $h$ is the Standard Model Higgs doublet and $v \approx 246$ GeV; in the MSSM context it may be thought of as the doublet that remains after integrating out the linear combination of $H_u$ and $H_d$ that does not obtain a VEV. The often-discussed $U$ parameter corresponds to a dimension-8 operator, $\left(h^\dagger\sigma^i h W^i_{\mu \nu}\right)^2$, and we can safely neglect it. In equating $S$ and $T$ with coefficients in ${\cal L}_{\rm oblique}$, we must first rewrite the Lagrangian (using equations of motion and integration by parts) in terms of a minimal basis of operators~\cite{Grzadkowski:2010es}. Other operators like $i \partial^\nu B_{\mu \nu} h^\dagger \overset{\leftrightarrow}{D^\mu} h$ will contribute to the $S$ parameter if we leave the result in terms of an overcomplete basis. We will see some examples below in which a straightforward diagrammatic calculation leads to operators not present in the minimal basis.

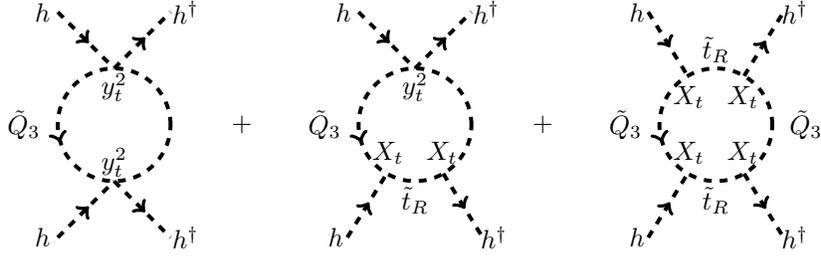
\begin{figure}[!h]\begin{center}
\begin{tikzpicture}[line width=1.5 pt]
\node at (-0.95,1.5) {$h$};
\node at (0.95,1.5) {$h^\dagger$};
\node at (-0.95,-1.5) {$h$};
\node at (0.95,-1.5) {$h^\dagger$};
\node at (-1.2, 0.0) {${\tilde Q}_3$};
\node at (0.0,0.5) {$y_t^2$};
\node at (0.0,-0.5) {$y_t^2$};
\draw[scalar] (-0.75,1.5)--(0.,0.75);
\draw[scalar] (0.,0.75)--(0.75,1.5);
\draw[scalar] (0,0) circle (0.75);
\draw[scalar] (-0.75,-1.5)--(0.,-0.75);
\draw[scalar] (0.,-0.75)--(0.75,-1.5);
\begin{scope}[shift={(4.0,0)}]
\node at (-2.3,0.0) {$+$};
\node at (-0.95,1.5) {$h$};
\node at (0.95,1.5) {$h^\dagger$};
\node at (-1.05,-1.5) {$h$};
\node at (1.05,-1.5) {$h^\dagger$};
\node at (-1.2, 0.0) {${\tilde Q}_3$};
\node at (0.0,0.5) {$y_t^2$};
\node at (-0.36,-0.37) {$X_t$};
\node at (0.36,-0.37) {$X_t$};
\node at (0.0,-1.0) {${\tilde t}_R$};
\draw[scalar] (-0.75,1.5)--(0.,0.75);
\draw[scalar] (0.,0.75)--(0.75,1.5);
\draw[scalar] (0,0) circle (0.75);
\draw[scalar] (-0.9,-1.5)--(-0.375,-0.6495);
\draw[scalar] (0.375,-0.6495)--(0.9,-1.5);
\end{scope}
\begin{scope}[shift={(8.0,0)}]
\node at (-2.3,0.0) {$+$};
\node at (-1.05,1.5) {$h$};
\node at (1.05,1.5) {$h^\dagger$};
\node at (-1.05,-1.5) {$h$};
\node at (1.05,-1.5) {$h^\dagger$};
\node at (-1.2, 0.0) {${\tilde Q}_3$};
\node at (1.2, 0.0) {${\tilde Q}_3$};
\node at (-0.36,-0.37) {$X_t$};
\node at (0.36,-0.37) {$X_t$};
\node at (-0.36,0.37) {$X_t$};
\node at (0.36,0.37) {$X_t$};
\node at (0.0,-1.0) {${\tilde t}_R$};
\node at (0.0,1.0) {${\tilde t}_R$};
\draw[scalar] (-0.9,1.5)--(-0.375,0.6495);
\draw[scalar] (0.375,0.6495)--(0.9,1.5);
\draw[scalar] (0,0) circle (0.75);
\draw[scalar] (-0.9,-1.5)--(-0.375,-0.6495);
\draw[scalar] (0.375,-0.6495)--(0.9,-1.5);
\end{scope}
\end{tikzpicture}
\end{center}
\caption{Loop diagrams contributing to the $T$ parameter operator $\left(h^\dagger D_\mu h\right)^2$ when the left-handed stop/sbottom doublet ${\tilde Q}_3$ and the right-handed stop ${\tilde t}_R = \left({\tilde u}^c_3\right)^\dagger$ are integrated out.}
\label{fig:Tdiagram}
\end{figure}

Integrating out any SU(2)$_L$ multiplet containing states that are split by electroweak symmetry breaking---for instance, the left-handed doublet of stops and sbottoms---will produce a contribution to $S$. The masses must additionally be split by custodial symmetry-violating effects to contribute to $T$. In the case of the stop and sbottom sector we have both, and $T$ is numerically dominant~\cite{Drees:1991zk}. The diagrams leading to a $T$-parameter are shown in Fig.~\ref{fig:Tdiagram}. There are terms proportional to $y_t^4$, to $y_t^2 X_t^2$, and to $X_t^4$. These diagrams are very familiar from the loop corrections to the Higgs quartic coupling that can lift the MSSM Higgs mass above the $Z$-mass~\cite{Haber:1990aw,Barbieri:1990ja,Casas:1994us,Carena:1995bx}. The only difference for $T$ is that we extract momentum-dependent terms to obtain the dimension-six operator. The result is:
\beq
T \approx \frac{m_t^4}{16 \pi \sin^2 \theta_W m_W^2 m_{{\tilde Q}_3}^2} + {\cal O}\left(\frac{m_t^2 X_t^2}{4 \pi m_{{\tilde Q}_3}^2 m_{{\tilde u}_3}^2}\right).
\eeq

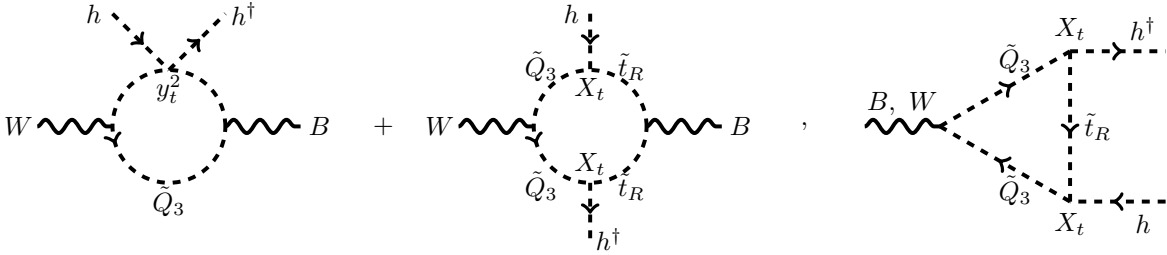
\begin{figure}[!h]\begin{center}
\begin{tikzpicture}[line width=1.5 pt]
\node at (-1.25,0.0) {$W$};
\node at (2.75,0.0) {$B$};
\draw[photon] (-1,0)--(0,0);
\draw[scalar] (0.75,0) circle (0.75);
\draw[photon] (1.5,0)--(2.5,0);
\draw[scalar] (0.0,1.5)--(0.75,0.75);
\draw[scalar] (0.75,0.75)--(1.5,1.5);
\node at (-0.25,1.5) {$h$};
\node at (1.75,1.5) {$h^\dagger$};
\node at (0.75,-1.0) {${\tilde Q}_3$};
\node at (0.75,0.5) {$y_t^2$};
\begin{scope}[shift={(5.6,0)}]
\node at (-2,0.0) {$+$};
\node at (-1.25,0.0) {$W$};
\node at (2.75,0.0) {$B$};
\draw[photon] (-1,0)--(0,0);
\draw[scalar] (0.75,0) circle (0.75);
\draw[photon] (1.5,0)--(2.5,0);
\draw[scalar] (0.75,1.5)--(0.75,0.75);
\draw[scalar] (0.75,-0.75)--(0.75,-1.5);
\node at (0.5,1.5) {$h$};
\node at (1.0,-1.5) {$h^\dagger$};
\node at (0.1,0.8) {${\tilde Q}_3$};
\node at (1.3,0.8) {${\tilde t}_R$};
\node at (0.1,-0.8) {${\tilde Q}_3$};
\node at (1.3,-0.8) {${\tilde t}_R$};
\node at (0.75,0.5) {$X_t$};
\node at (0.75,-0.5) {$X_t$};
\end{scope}
\begin{scope}[shift={(11.0,0)}]
\node at (-1.8,0) {$,$};
\node at (-0.5,0.3) {$B,~W$};
\node at (1.0,0.9) {${\tilde Q}_3$};
\node at (1.0,-0.85) {${\tilde Q}_3$};
\node at (2.1,0.0) {${\tilde t}_R$};
\node at (2.7,1.3) {$h^\dagger$};
\node at (2.7,-1.3) {$h$};
\node at (1.73,1.3) {$X_t$};
\node at (1.73,-1.3) {$X_t$};
\draw[photon] (-1,0)--(0,0);
\draw[scalar] (0,0)--(1.73,1.0);
\draw[scalar] (1.73,1.0)--(1.73,-1.0);
\draw[scalar] (1.73,-1.0)--(0,0);
\draw[scalar] (1.73,1.0)--(3.0,1.0);
\draw[scalar] (3.0,-1.0)--(1.73,-1.0);
\end{scope}
\end{tikzpicture}
\end{center}
\caption{Loop diagrams contributing to the $S$ parameter. The two diagrams at left generate the usual operator $h^\dagger W^{i \mu \nu} \sigma^i h B_{\mu \nu}$ when the left-handed stop/sbottom doublet ${\tilde Q}_3$ and the right-handed stop ${\tilde t}_R = \left({\tilde u}^c_3\right)^\dagger$ are integrated out. The diagram at right generates the operators $i \partial^\nu B_{\mu \nu} h^\dagger \overset{\leftrightarrow}{D^\mu} h$ and $i D^\nu W^i_{\mu \nu} h^\dagger  \sigma^i \overset{\leftrightarrow}{D^\mu} h$, which also contribute to $S$ after being rewritten in terms of the minimal basis of dimension-six operators.}
\label{fig:Sdiagram}
\end{figure}

The diagrams generating the $S$-parameter are shown in Fig.~\ref{fig:Sdiagram}. Notice that in order for the first diagram to contribute, it is important that the SU(2)$_L$ structure of the coupling is $\left(h \cdot {\tilde Q}_3\right)\left(h^\dagger \cdot {\tilde Q}_3^\dagger\right)$ rather than $(h^\dagger h)({\tilde Q}_3^\dagger {\tilde Q}_3)$, as the latter would lead to a zero SU(2)$_L$ trace around the loop. As a result, the $F$-term potential contributes $\propto y_t^2$ and the SU(2)$_L$ $D$-term potential contributes $\propto g^2$, but there is no U(1)$_Y$ $D$-term contribution $\propto g'^2$. The leading correction is
\beq
S \approx -\frac{1}{6 \pi} \frac{m_t^2}{m_{{\tilde Q}_3}^2} + {\cal O}\left(\frac{m_t^2 X_t^2}{4 \pi m_{{\tilde Q}_3}^2 m_{{\tilde u}_3}^2}\right).
\eeq
The $X_t$ dependent part of the correction depends on the subtlety in the use of our effective oblique Lagrangian eq.~\ref{eq:Loblique} that we mentioned above: the strict relation between $S$ and the coefficient of $h^\dagger W^{i \mu \nu} \sigma^i h B_{\mu \nu}$ applies only if we first rewrite all operators in a minimal basis~\cite{Grojean:2006nn,Han:2008es}. The third loop diagram of Fig.~\ref{fig:Sdiagram} generates different operators like $i \partial^\nu B_{\mu \nu} h^\dagger \overset{\leftrightarrow}{D^\mu} h$ which may be rewritten using integration by parts and equations of motion and also contribute to $S$. Note that a similar diagram with a bubble topology connecting a gauge boson on one side and two Higgs bosons on the other (which can be obtained by removing one of the vector bosons from the left most diagram in Fig.~\ref{fig:Sdiagram}) cannot be sensitive to the {\em difference} in momenta of the Higgs bosons, and so never generates the operators in question. The fact that integrating out heavy particles often generates operators that are not present in the minimal basis was also recently emphasized in ref.~\cite{Henning:2014gca, Henning:2014wua}.

Notice that the $S$ parameter contribution from loops of stops and sbottoms is small and, for small $X_t$, negative. The $T$ parameter contribution is numerically somewhat larger and positive. In both cases, the dominant contribution is due to the left-handed stops and sbottoms, with their right-handed counterparts entering through mixing effects. As a result, we expect that precision measurements of the $T$ parameter can set interesting constraints on left-handed stops. (For a recent study of existing constraints, see ref.~\cite{Espinosa:2012in}.)

\subsection{Production of $b$ and $t$ Quarks}
\label{subsec:Rbops}

Integrating out loops of stops and higgsinos can correct the production of bottom and top quarks at $e^+ e^-$ colliders. In particular, in the minimal basis of dimension-six operators these corrections show up in the terms~\cite{Grzadkowski:2010es}
\beq
%{\cal L}_{bt} = 
c^h_{q;1} i h^\dagger \overset{\leftrightarrow}{D_\mu} h Q_3^\dagger {\overline \sigma}^\mu Q_3 + c^h_{q;3} i h^\dagger \sigma^i\overset{\leftrightarrow}{D_\mu} h Q_3^\dagger \sigma^i {\overline \sigma}^\mu Q_3 + c^h_{u} i h^\dagger \overset{\leftrightarrow}{D_\mu} h u^{c\dagger}_3 {\overline \sigma}^\mu u^c_3 + c^h_{d} i h^\dagger \overset{\leftrightarrow}{D_\mu} h d^{c\dagger}_3 {\overline \sigma}^\mu d^c_3 + {\rm h.c.}
\label{eq:Lbt}
\eeq
Again, however, calculating loop diagrams might generate other operators not present in Eq.~\ref{eq:Lbt}, in which case we should use the equations of motion and integration by parts to rewrite the operators in a minimal basis.

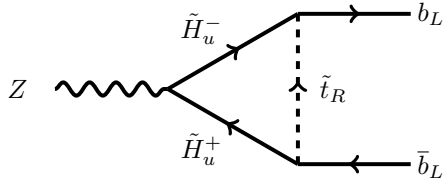
\begin{figure}[!h]\begin{center}
\begin{tikzpicture}[line width=1.5 pt]
\draw[photon] (-0.5,0)--(1,0);
\draw[electron] (1,0)--(2.73,1.0);
\draw[electron] (2.73,-1.0)--(1,0);
\draw[scalar] (2.73,-1)--(2.73,1);
\draw[electron] (2.73,1)--(4.23,1.0);
\draw[electron] (4.23,-1.0)--(2.73,-1);
\node at (-1.0,0) {$Z$};
\node at (4.5,1.0) {$b_L$};
\node at (4.5,-1.0) {$\overline{b}_L$};
\node at (3.2,0.0) {${\tilde t}_R$};
\node at (1.45,0.8) {$\tilde{H}_u^-$};
\node at (1.45,-0.8) {$\tilde{H}_u^+$};
\end{tikzpicture}
\end{center}
\caption{Loop diagram correcting $R_b$. The operators that are generated are $W^i_{\mu \nu} Q_3^\dagger \sigma^i \overline{\sigma}^\mu i D^\nu Q_3$ and $B_{\mu \nu} Q_3^\dagger \overline{\sigma}^\mu i D^\nu Q_3$. }
\label{fig:Zbbcorrection}
\end{figure}

The largest effects are associated with the top quark Yukawa coupling $y_t u^c_3 H_u \cdot Q_3$. As a result, we should look for corrections associated with the production of {\em left}-handed $b$ quarks, and either left- or right-handed top quarks. Let us begin by discussing the $b$-quark coupling, which is constrained for instance by measurements of
\beq
R_b \equiv \frac{\Gamma(Z \to b{\overline b})}{\Gamma(Z \to {\rm hadrons})}.
\eeq
A diagram generating a correction to the $Z \to b{\overline b}$ process is shown in Fig.~\ref{fig:Zbbcorrection}. This cannot arise from an operator in eq.~\ref{eq:Lbt}, because there is nowhere in the diagram that we could place insertions of $h$ and $h^\dagger$. A more complete list of operators~\cite{Buchmuller:1985jz} includes the additional terms
\beq
W^i_{\mu \nu} Q_3^\dagger \sigma^i \overline{\sigma}^\mu i D^\nu Q_3,~~B_{\mu \nu} Q_3^\dagger \overline{\sigma}^\mu i D^\nu Q_3,
\eeq
which also couple the left-handed bottom quark to the $Z$ boson. These operators, missing in the minimal basis, are the ones that are generated by integrating out higgsinos and right-handed stops. (Note the similarity in form of both the diagram and the corresponding operator to the right-hand diagram of fig.~\ref{fig:Sdiagram}.) The full dependence of $R_b$ on dimension-six operators is worked out in ref.~\cite{Oakes:1999zi}.

In fact, we can understand the expected size of the resulting effect in somewhat more detail by integrating out {\em first} the right-handed stops and subsequently the higgsinos. After the first step we have a four-fermion operator: 
\beq
\begin{tikzpicture}[line width=1.5 pt]
\draw[electron] (0,0)--(1,0);
\node at (0.5,0.5) {${\tilde H}_u$};
\draw[electron] (1,0)--(2,0);
\node at (1.5,0.5) {$b_L$};
\draw[scalar] (1,-1)--(1,0);
\node at (1.5,-0.5) {${\tilde t}_R$};
\draw[electron] (1,-1)--(0,-1);
\node at (0.5, -1.5) {${\tilde H}_u$};
\draw[electron] (2,-1)--(1,-1);
\node at (1.5,-1.5) {$\overline{b}_L$};
\node at (4.5,-0.5) {\large $\Rightarrow \frac{y_t^2}{m_{{\tilde t}_R^2}}\left({\tilde H}_u \cdot Q_3\right)\left(Q_3^\dagger \cdot {\tilde H}_u^\dagger\right).$};
\end{tikzpicture}
\eeq
This operator then mixes with the $Zb\overline{b}$ coupling as we integrate out the higgsinos:
\beq
\begin{tikzpicture}[line width=1.5 pt]
\draw[photon] (0,0)--(1,0);
\node at (-0.3,0) {$Z$};
\draw[electron] (1.75,0) circle (0.75);
\node at (1.5,1.0) {${\tilde H}_u$};
\draw[electron] (2.5,0)--(3.2,0.7);
\draw[electron] (3.2,-0.7)--(2.5,0);
\node at (3.4,0.7) {$b_L$};
\node at (3.4,-0.7) {$\overline{b}_L$};
\node at (6.5,0.0) {\large $\Rightarrow \frac{y_t^2}{m_{{\tilde t}_R^2}} W^i_{\mu \nu} Q_3^\dagger \sigma^i \overline{\sigma}^\mu i D^\nu Q_3~\log \frac{m_{{\tilde t}_R}}{\mu}.$};
\end{tikzpicture}
\label{eq:Rbnomixing}
\eeq
The structure of derivatives in this operator produces a factor of $m_Z^2$ in the formula for $R_b$, eq.~\ref{eq:deltaRbformula}. The reason for integrating the particles out in two steps is to highlight that there is a potentially large logarithm of the ratio of stop and higgsino masses. In a careful effective field theory treatment, this log could be resummed by computing the renormalization group evolution that mixes the four-fermion operator with the operator modifying the $Z$ coupling through their matrix of anomalous dimensions.

Once we include mixing of the left- and right-handed stops, there are additional terms that directly generate the operators in eq.~\ref{eq:Lbt}. We can start by integrating out the left-handed stops to generate a correction to the coupling of right-handed stops to the $Z$ boson:
\beq
\begin{tikzpicture}[line width=1.5 pt]
\draw[photon] (0,0)--(1.5,0);
\draw[scalar] (1.5,0)--(2.25,0.75);
\draw[scalar] (2.25,0.75)--(3.0,1.5);
\draw[scalar] (3.0,-1.5)--(2.25,-0.75);
\draw[scalar] (2.25,-0.75)--(1.5,0);
\draw[dashed] (2.25,0.75)--(1.5,1.5);
\draw[dashed] (2.25,-0.75)--(1.5,-1.5);
\node at (-0.375,0) {$Z$};
\node at (1.2,1.5) {$h$};
\node at (1.2,-1.5) {$h$};
\node at (1.55,0.55) {${\tilde t}_L$};
\node at (1.6,-0.55) {${\tilde t}_L$};
\node at (2.6,0.75) {${\tilde t}_R$};
\node at (2.6,-0.75) {${\tilde t}_R$};
\node at (5.2,0.0) {\Large $ \Rightarrow \frac{y_t^2 X_t^2 \left(h^\dagger i \overleftrightarrow{D}_\mu h\right)\left({\tilde t}_R^\dagger i \overleftrightarrow{D}^\mu {\tilde t}_R\right)}{m_{{\tilde t}_L}^4}.$};
\end{tikzpicture}
\eeq
This new operator then mixes at one loop into the operator coupling $Z$ bosons to the left-handed $b$ quark:
\beq
\begin{tikzpicture}[line width=1.5 pt]
\draw[photon] (0,0)--(1,0);
\node at (-0.3,0) {$Z$};
\draw[dashed] (1,0)--(0.5,0.5);
\node at (0.2,0.5) {$h$};
\draw[dashed] (1,0)--(0.5,-0.5);
\node at (0.2,-0.5) {$h$};
\draw[scalar] (1,0)--(1.7,0.7);
\draw[scalar] (1.7,-0.7)--(1,0);
\draw (1.7,0.7)--(1.7,-0.7);
\node at (2.0,0.0) {${\tilde H}_u$};
\draw[electron] (1.7,0.7)--(2.4,0.7);
\node at (2.7,0.7) {$b_L$};
\draw[electron] (2.4,-0.7)--(1.7,-0.7);
\node at (2.7,-0.7) {$\overline{b}_L$};
\node at (7.0,0.0) {\Large $ \Rightarrow \frac{y_t^4 X_t^2 \left(h^\dagger i \overleftrightarrow{D}_\mu h\right)\left(Q_3^\dagger \overline{\sigma}^\mu Q_3\right)}{m_{{\tilde t}_L}^4} \log \frac{m_{{\tilde t}_L}}{{\rm max}(m_{{\tilde t}_R}, \mu)}.$};
\end{tikzpicture}
\label{eq:Rbmixing}
\eeq
These structures that we have deduced on effective field theory grounds match terms that can be found by expanding the full loop formulas in refs.~\cite{Boulware:1991vp, Wells:1994cu}.

A future $e^+ e^-$ collider running above the $t{\overline t}$ threshold can also measure corrections to the top quark's couplings to $Z$ bosons and photons to about 1\% accuracy~\cite{Amjad:2013tlv,Asner:2013hla}. The $Zt_L{\overline t}_L$ vertex is modified by the same operator as $R_b$, and a correction to the $Zt_R {\overline t}_R$ vertex can also arise from integrating out left-handed stops. We expect that either $R_b$ or the $T$ parameter will provide stronger constraints in any region of parameter space that modifies the $t{\overline t}$ couplings, though depending on the details of a future collider and the luminosity it accumulates for top quark production this may need to be revisited in the future.

\subsection{Higgs Couplings to Photons and Gluons}
\label{sec:higgsgammaglue}

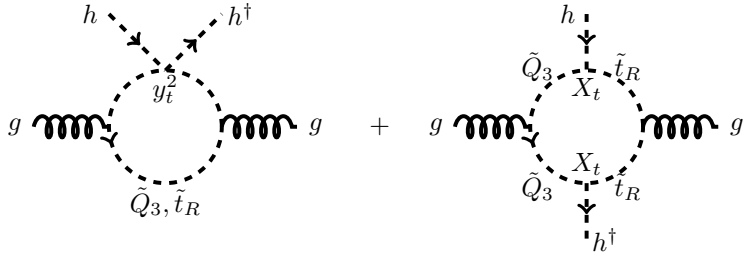
\begin{figure}[!h]\begin{center}
\begin{tikzpicture}[line width=1.5 pt]
\node at (-1.25,0.0) {$g$};
\node at (2.75,0.0) {$g$};
\draw[gluon] (-1,0)--(0,0);
\draw[scalar] (0.75,0) circle (0.75);
\draw[gluon] (1.5,0)--(2.5,0);
\draw[scalar] (0.0,1.5)--(0.75,0.75);
\draw[scalar] (0.75,0.75)--(1.5,1.5);
\node at (-0.25,1.5) {$h$};
\node at (1.75,1.5) {$h^\dagger$};
\node at (0.75,-1.0) {${\tilde Q}_3, {\tilde t}_R$};
\node at (0.75,0.5) {$y_t^2$};
\begin{scope}[shift={(5.6,0)}]
\node at (-2,0.0) {$+$};
\node at (-1.25,0.0) {$g$};
\node at (2.75,0.0) {$g$};
\draw[gluon] (-1,0)--(0,0);
\draw[scalar] (0.75,0) circle (0.75);
\draw[gluon] (1.5,0)--(2.5,0);
\draw[scalar] (0.75,1.5)--(0.75,0.75);
\draw[scalar] (0.75,-0.75)--(0.75,-1.5);
\node at (0.5,1.5) {$h$};
\node at (1.0,-1.5) {$h^\dagger$};
\node at (0.1,0.8) {${\tilde Q}_3$};
\node at (1.3,0.8) {${\tilde t}_R$};
\node at (0.1,-0.8) {${\tilde Q}_3$};
\node at (1.3,-0.8) {${\tilde t}_R$};
\node at (0.75,0.5) {$X_t$};
\node at (0.75,-0.5) {$X_t$};
\end{scope}
\end{tikzpicture}
\end{center}
\caption{Loop diagrams contributing to the correction to the Higgs coupling to gluons, via the operator $h^\dagger h G^a_{\mu \nu} G^{a \mu \nu}$.}
\label{fig:higgscoupling}
\end{figure}%

The corrections to the Higgs couplings induced by loops of stops and sbottoms have been the subject of intense recent interest~\cite{Carena:2011aa,Arvanitaki:2011ck,Blum:2012ii,Espinosa:2012in,Kribs:2013lua,Fan:2014txa}. As is well known, stop loops could modify the Higgs coupling to gluons, via diagrams like those of Fig.~\ref{fig:higgscoupling}. The leading order contribution could be computed easily via the low energy Higgs theorem~\cite{Ellis:1975ap, Shifman:1979eb}
\beq\label{eq:rG}
r_G^{\tilde t} \equiv \frac{c_{hgg}^{\tilde t}}{c_{hgg}^{\rm SM}}
\approx \frac{1}{4} \left(\frac{m_t^2}{m_{\tilde{t}_1}^2}+\frac{m_t^2}{m_{\tilde{t}_2}^2}-\frac{m_t^2X_t^2}{m_{\tilde{t}_1}^2m_{\tilde{t}_2}^2}\right), \quad {\rm stop~contribution~to~} hgg ~{\rm coupling}
\label{eq:hgg}
\eeq
where we neglect $D$-terms. The low-energy theorem essentially upgrades the $\log(M_{\rm threshold})$ terms that appear when integrating out a heavy mass threshold to field-dependent terms, viewing $M_{\rm threshold}$ as a function of a variable higgs VEV. The resulting expression is valid for $m_{{\tilde t}_{1,2}} \gsim m_h/2$, which we will assume is always true. A loop of light stops will also generate a smaller contribution to the Higgs diphoton coupling, which is anti-correlated to $r_G^{\tilde t}$
\beq\label{eq:gamG} r_\gamma^{\tilde t}  \equiv \frac{c_{h\gamma\gamma}^{\tilde t}}{c_{h\gamma\gamma}^{\rm SM}}=\frac{\mathcal{A}^\gamma_{\tilde t}}{\left(\mathcal{A}^\gamma_W+\mathcal{A}^\gamma_t\right)^{\rm SM}}
\approx -0.28 r_G^{\tilde t},\eeq
using $\mathcal{A}^\gamma_W\approx8.33$ and $\mathcal{A}^\gamma_t\approx-1.84$, the amplitudes of $h \to \gamma\gamma$ in the SM, valid for $m_h=125$ GeV. 
One could see that the more natural the stop parameter space is, the larger the modification is~\cite{Blum:2012ii}. Except for the special case of colorless stop, the strongest limit on the stop always comes from the measurement of $hgg$ coupling.

Corrections to $\Gamma(h \to Z \gamma)$ play a similar role as those for $\Gamma(h \to \gamma\gamma)$, but we find that they are numerically less important. Similarly, corrections to the Higgs coupling to $Z$ bosons play a subdominant role because they compete with the large tree-level coupling.

\subsection{Wavefunction Renormalization}
\label{sec:wavefunction}

Recently ref.~\cite{Craig:2013xia} has emphasized that any new physics which couples to the Higgs will induce a wavefunction renormalization of the Higgs boson, arising from the dimension-six kinetic term $\partial_\mu \left|h\right|^2 \partial^\mu \left|h\right|^2$  (also see~\cite{Englert:2013tya, Gori:2013mia}). This is an interesting observation, because it opens up the possibility of probing naturalness even in scenarios where the quadratic divergence in the Higgs mass is canceled by particles without Standard Model quantum numbers, which are otherwise hard to probe. We have generalized the calculation of this correction from ref.~\cite{Englert:2013tya} to allow for mixing between the two stops. We write the general result in terms of the couplings of stop mass eigenstates to the Higgs boson, $g_{hij} \equiv g_{h{\tilde t}^\dagger_i {\tilde t}_j}$:
\beq
g_{h11} & = & y_t^2 v - \frac{y_t X_t}{\sqrt{2}} \sin(2\theta_{\tilde t}) = y_t^2 v \left(1 - \frac{X_t^2}{m_{{\tilde t}_2}^2 - m_{{\tilde t}_1}^2}\right), \nonumber \\
g_{h22} & = & y_t^2 v + \frac{y_t X_t}{\sqrt{2}} \sin(2\theta_{\tilde t}) = y_t^2 v \left(1 + \frac{X_t^2}{m_{{\tilde t}_2}^2 - m_{{\tilde t}_1}^2}\right), \nonumber \\
g_{h12} = g_{h21} & = & \frac{y_t X_t}{\sqrt{2}} \cos(2\theta_{\tilde t}).
\label{eq:stopstophiggs}
\eeq
In terms of these couplings, the fractional correction to the Higgsstrahlung cross section is given by
\beq
\delta \sigma_{Zh} = \sum_{i = 1}^2 \sum_{j=1}^2 \frac{N_c g_{hij}^2}{16 \pi^2} I\left(m_h^2; m_{{\tilde t}_i}^2, m_{{\tilde t}_j}^2\right), 
\eeq
where the loop function is
\beq
I\left(p^2; m_1^2, m_2^2\right)  = \int_0^1 dx~\frac{x(1-x)}{x(1-x)p^2 - x m_1^2 - (1-x) m_2^2}.
\eeq
In the special case of equal masses, this is 
\beq
I\left(p^2; m^2, m^2\right) = \frac{1}{p^2} \left(1 - \frac{4 m^2~{\rm arctan}\left(\sqrt{\frac{p^2}{4 m^2 - p^2}}\right)}{\sqrt{p^2 (4 m^2 - p^2)}}\right),
\eeq
in agreement with the result of ref.~\cite{Englert:2013tya}. For us the more relevant limit is the case $m_h^2 \ll m_{{\tilde t}_1}^2,m_{{\tilde t}_2}^2$, in which case the loop function reduces to: 
\beq
I(0; m_{\tilde{t}_1}^2,m_{\tilde{t}_2}^2) = \frac{m_{\tilde{t}_1}^2 m_{\tilde{t}_2}^2 \log(m_{\tilde{t}_1}^2/m_{\tilde{t}_2}^2)}{\left(m_{\tilde{t}_1}^2-m_{\tilde{t}_2}^2\right)^3} - \frac{m_{\tilde{t}_1}^2 + m_{\tilde{t}_2}^2}{2\left(m_{\tilde{t}_1}^2-m_{\tilde{t}_2}^2\right)^2} \to -\frac{1}{6 m^2}~{\rm if}~m_{\tilde{t}_1}^2=m_{\tilde{t}_1}^2=m^2.
\eeq
This scaling is as expected from the interpretation of the wavefunction renormalization as the coefficient of the dimension-six operator $\partial_\mu \left|h\right|^2 \partial^\mu \left|h\right|^2$.

The approximate FCC-ee measurement limit, $\delta \sigma_{Zh} \approx 0.2\%$, is reached when $X_t = 0$ for equal stop masses of 440 GeV. However, as computed in detail recently in ref.~\cite{Craig:2014una}, several different operators contribute to the Higgsstrahlung cross section when stops are integrated out. The wavefunction renormalization is one contribution, but others arise from operators like $h^\dagger h W^i_{\mu \nu} W^{i \mu \nu}$ that directly alter the Higgs coupling to the $Z$ boson. As shown in figure 3 of their paper, the effect of including these additional operators on the Higgsstrahlung cross section is roughly a factor of 2 larger than including the wavefunction correction alone (with some dependence on the stop masses---light left-handed stops play a larger role than light right-handed stops). As a result, the bounds can be slightly larger than those estimated from wavefunction renormalization alone. Nonetheless, the future $e^+ e^-$ collider reach from measurements of $\Gamma(h \to gg)$, as estimated in refs.~\cite{Fan:2014txa,Henning:2014gca,Craig:2014una,Fan} and in this paper below, is more constraining than that from Higgsstrahlung while the measurement of $\Gamma(h \to \gamma\gamma)$ (combined with the HL-LHC result) is comparable to or a bit weaker than those from Higgsstrahlung.

\subsection{Other Corrections to Precision Observables}
\label{sec:othercorrections}

Loops of stops and higgsinos are likely to give the dominant correction to the $b \to s \gamma$ amplitude in natural SUSY theories~\cite{Barbieri:1993av,Okada:1993sx,Ishiwata:2011ab,Blum:2012ii,Espinosa:2012in,Altmannshofer:2012ks,Katz:2014mba}. Although this is an interesting bound on natural SUSY parameter space, it depends on a combination $A_t \mu \tan \beta/m_{{\tilde t}}^4$, and so results in a weaker constraint on $A_t$ when $\tan \beta$ is small. This has interesting implications for the heavy Higgs bosons of the 2HDM, $H^0$, $A^0$, and $H^\pm$, which should not be too heavy~\cite{Gherghetta:2014xea,Katz:2014mba} and may have interesting effects of their own on precision observables~\cite{Blum:2012ii,Gupta:2012fy}. As we will discuss in Sec. 7.1, it could be the main sensitive probe to the ``blind spot" region. 

Charginos and neutralinos have relatively small effects on the observables we have mentioned so far. This is largely because they have dominantly vectorlike masses and sensitivity to SU(2)$_L$ breaking through the Higgs is a small effect. On the other hand, integrating out higgsinos or winos will always generate the triple gauge coupling operator $c_{WWW} g \epsilon_{ijk} W^i_{\mu \nu} W^{j\nu}_\rho W^{k\rho\mu}$. Unfortunately, the coefficient generated by integrating out an SU(2)$_L$ multiplet is small~\cite{Cho:1994yu}:
\beq
c_{WWW} = \frac{g^2}{2880 \pi^2} \sum_{{\rm rep}~R,~{\rm mass}~M} \left(-1\right)^F \frac{T(R)}{M^2},
\label{eq:cWWW}
\eeq
where $T(R)$ is the Dynkin index of the representation and the sum is over Weyl fermions for which $F = 1$ and complex scalars for which $F = 0$. (That the effect of a complex scalar and that of a Weyl fermion cancel for equal masses is a result of a supersymmetric Ward identity~\cite{Grisaru:1977px}.) Expected bounds from the ILC are expressed in terms of dimensionless coefficients $\lambda_\gamma$ and $\lambda_Z$, which are both equal to $6 m_W^2 c_{WWW}$. The ILC can bound the coefficient at 1$\sigma$ to be $\left|\lambda_{\gamma,Z}\right| \lsim 6 \times 10^{-4}$ with 500 fb$^{-1}$ at $\sqrt{s} = 500$ TeV or half that with 1 ab$^{-1}$ at $\sqrt{s} = 800$ GeV~\cite{AguilarSaavedra:2001rg,Baer:2013cma}. Even for the bound assuming higher energy and luminosity, this does not probe wino or higgsino (or left-handed stop) masses above 100 GeV.

Similarly, any particles with SU(2)$_L$ quantum numbers contribute above threshold to the running of gauge couplings. At future very high energy proton--proton colliders this might be detected with precision Drell-Yan measurements~\cite{AlvesToAppear}. At an $e^+ e^-$ collider it would be difficult, but if the collider attains high luminosities at energies near 1 TeV it may be possible to probe running. There is also a ``below-threshold running effect'' arising from the operator $c_{JJ} D^\mu W^i_{\mu \nu} D_\lambda W^{i\lambda \nu}$, which has coefficient~\cite{Cho:1994yu} 
\beq
c_{JJ} = -\frac{g^2}{960 \pi^2} \sum_{{\rm rep}~R,~{\rm mass}~M} a_F \frac{T(R)}{M^2},
\eeq
where $a_F = 4$ for Weyl fermions and $1$ for complex scalars. By the equation of motion, $D_\mu W^{i \mu \nu} = -g J^{i\nu}$, where $J^{i\nu}$ is the SU(2)$_L$ current, so this operator is a current--current interaction that may be thought of as a power-law ($p^2/M^2$) running of the gauge coupling below the scale $M$. In the usual QED calculation of vacuum polarization, one obtains an expression like $\int^1_0 dx~x(1-x) \log(M^2 - p^2 x (1-x))$ and expands for $-p^2 \gg M^2$ to obtain logarithmic running. This operator is simply the corresponding result if we expand for $M^2 \gg p^2$. Again, it will be difficult to obtain interesting constraints from this operator simply because the number in the denominator is so large.

\subsection{Comments on the Use of Effective Field Theory}

In the remainder of the paper we will use formulas for $S$, $T$, and $R_b$ originating in refs.~\cite{Drees:1991zk,Boulware:1991vp} and presented in Appendix~\ref{app:loops}. These include complete loop functions based on the original Peskin-Takeuchi definitions of $S$ and $T$ in terms of gauge boson vacuum polarizations, allowing for arbitrary stop-sector mixing. In particular, nontrivial functions of ratios like $m_t X_t/m_{{\tilde u}_3}^2$, if expanded in powers of the Higgs VEV, may effectively come from operators of dimension higher than 6 in an EFT treatment. In this sense, the full loop functions include effects of higher order than the operator analysis we have sketched above. On the other hand, as we have discussed in the case of $R_b$, if we integrate out multiple particles---say, first right-handed stops and then higgsinos---these loop functions may contain logarithms like $\log(m_{{\tilde t}_R}/\mu)$ that could be resummed. In this case a careful EFT operator analysis would first obtain a Wilson coefficient for an operator like $\left({\tilde H}_u \cdot Q_3\right)\left(Q_3^\dagger \cdot {\tilde H}_u^\dagger\right)$ and then use the matrix of anomalous dimensions in an EFT including the Standard Model fields and higgsinos to compute how this operator mixes through RG evolution into the operator $W^i_{\mu \nu} Q_3^\dagger \sigma^i {\overline \sigma}^\mu i D^\nu Q_3$ that modifies the $Zb_L {\overline b}_L$ coupling measured at low energy. The RG calculation would resum logarithms that are not resummed in the classic results of refs.~\cite{Drees:1991zk,Boulware:1991vp}.

For the purposes of setting limits, the detailed resummation probably produces only mild changes to the result, and so for now we use the loop formulas and forego the RG treatment. On the other hand, if one observes a deviation from SM expectations at a future collider, the RG calculation should be done to assess what parameters are preferred by the observations.

%%%%%%%%%%%%%%%%%%%%%%%%%%%%%%%%%%
\section{Oblique Corrections from Stop Sector}
\label{sec:oblique}
%%%%%%%%%%%%%%%%%%%%%%%%%%%%%%%%%%

%%%%%%%%%%%%%%%%%%%%%%%%%%%%%%%%%%
\subsection{Global Fit of Electroweak Observables with Oblique Corrections}
\label{sec:fitbasic}
%%%%%%%%%%%%%%%%%%%%%%%%%%%%%%%%%%

In our previous paper~\cite{Fan}, we performed a simplified global fit of electroweak observables with oblique corrections for current and future electroweak precision tests. The simplified fit includes five Standard Model observables that are free to be varied in the fit: the top quark mass $m_t$, the $Z$ boson mass $m_Z$, the Higgs boson mass $m_h$, the strong coupling constant at the $Z$ pole $\alpha_s(M_Z^2)$ and the hadronic contribution to the running of the fine-structure constant $\alpha$: $\Delta\alpha_{\rm had}^{(5)}(M_Z^2)$. It also includes three additional observables, the $W$ boson mass $m_W$, the weak mixing angle $\sin^2 \theta_{\rm eff}^{\ell}$ and the total $Z$ boson decay width $\Gamma_Z$, which are determined by the values of the five free observables in the Standard Model. The effects of the new physics are parametrized by the oblique parameters $S$ and $T$~\cite{Peskin:1991sw, Peskin:1990zt}. We constructed a modified $\chi^2$ function taking into account of the theory uncertainties with a flat prior and then performed profile likelihood fits to carve out the allowed $(S, T)$ regions for different experiments. We will refer the readers to~\cite{Fan} for details of the fit and the results. A quick summary of the results is that the current 1$\sigma$ allowed range of $S$ and $T$ is about 0.1 which will be reduced to $\lesssim 0.03$ at ILC and CEPC baseline plan, $\lesssim 0.01$ at FCC-ee and CEPC with potential improvements. The possible improvements of the CEPC electroweak program are described in~\cite{Fan}.

%%%%%%%%%%%%%%%%%%%%%%%%%%%%%%%%%%
\subsection{Constraints on the Stop Sector}
\label{sec:stopoblique}
%%%%%%%%%%%%%%%%%%%%%%%%%%%%%%%%%%

%%%%%%%%%%%%%%%%%%%%%%%%%%%%%%%%%%%%%%%%%%%%%%%%%%%%%%
\begin{figure}[!h]\begin{center}
\includegraphics[width=0.3\textwidth]{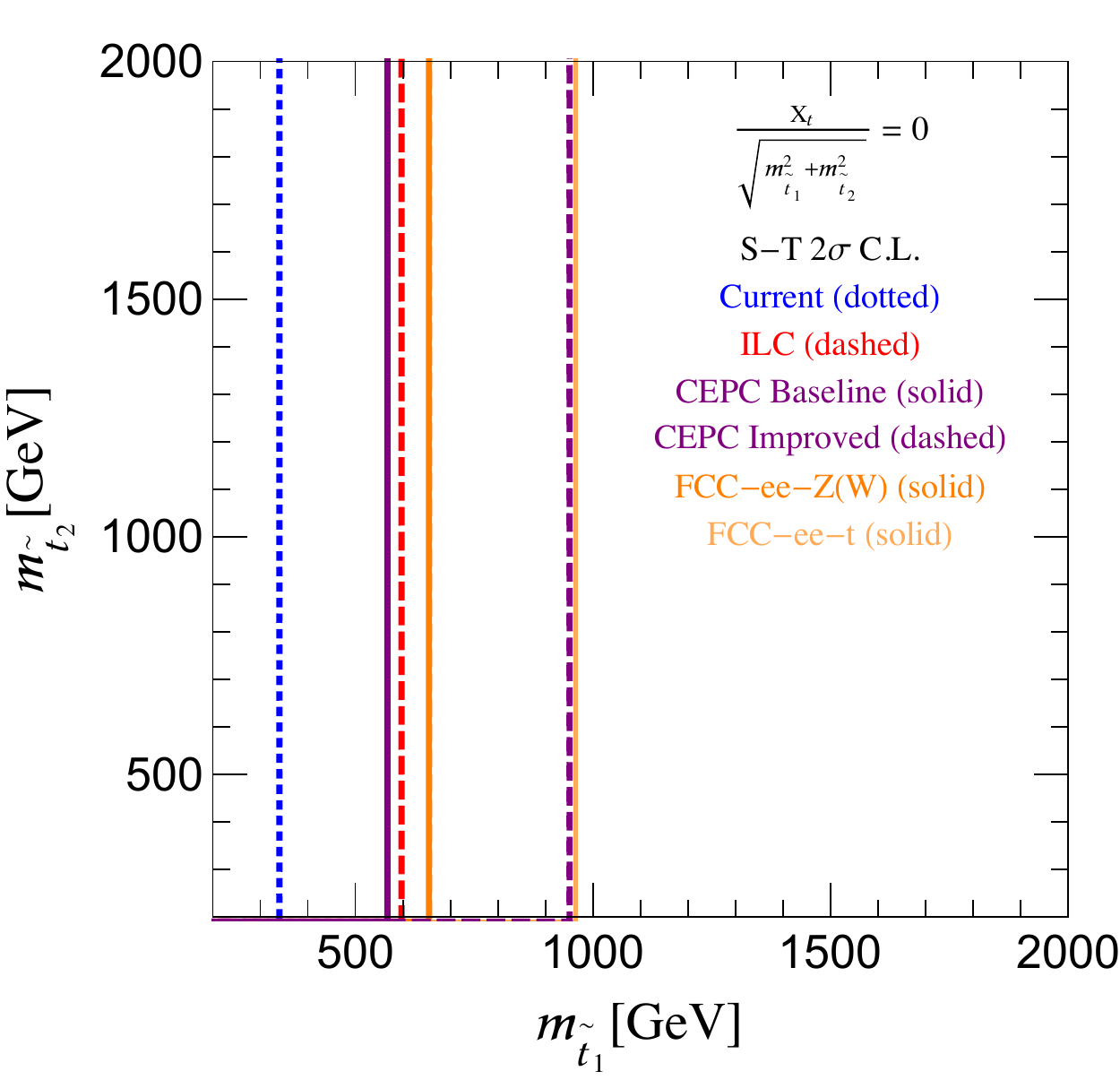} \quad  \includegraphics[width=0.3\textwidth]{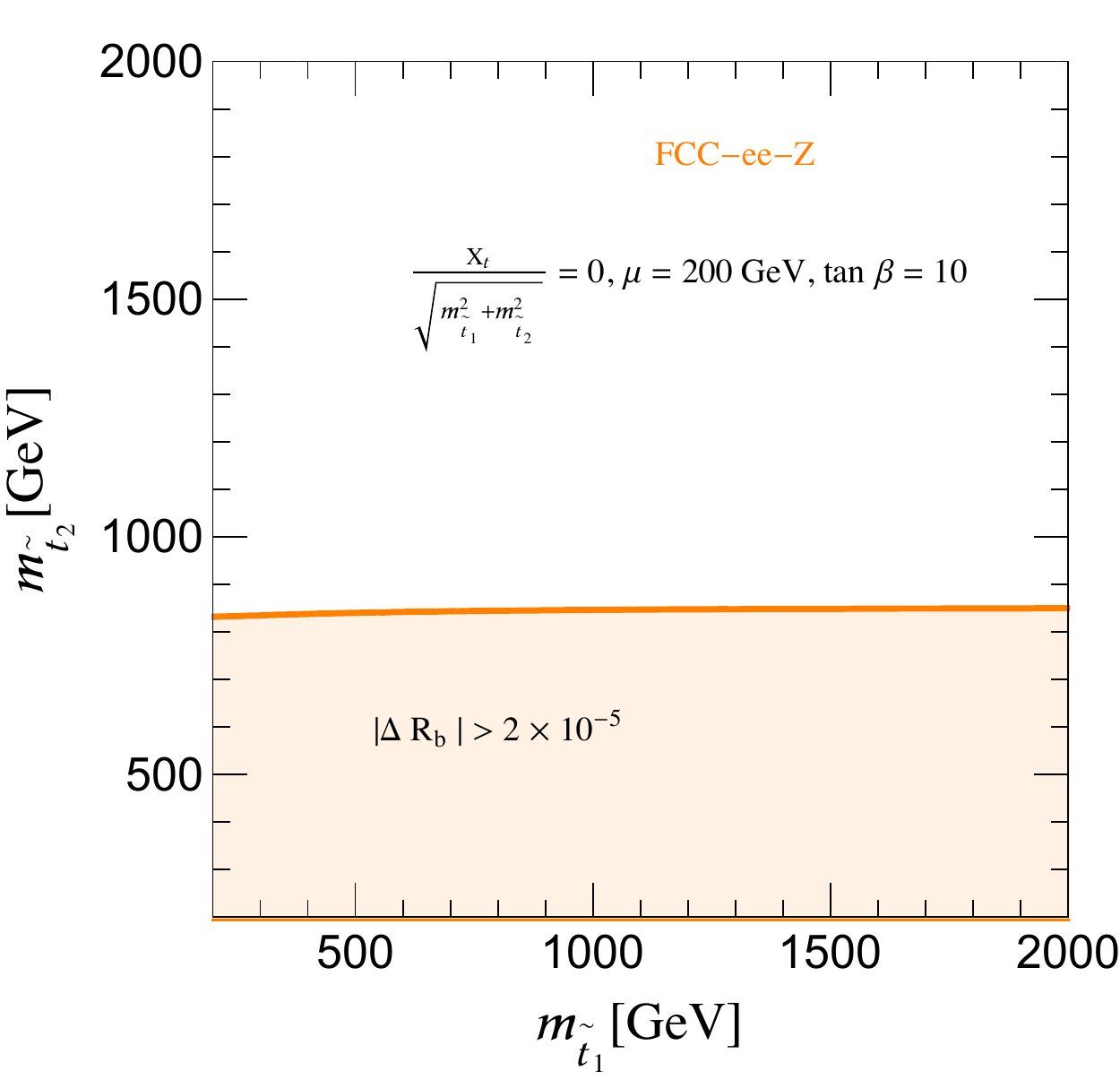} \quad  \includegraphics[width=0.3\textwidth]{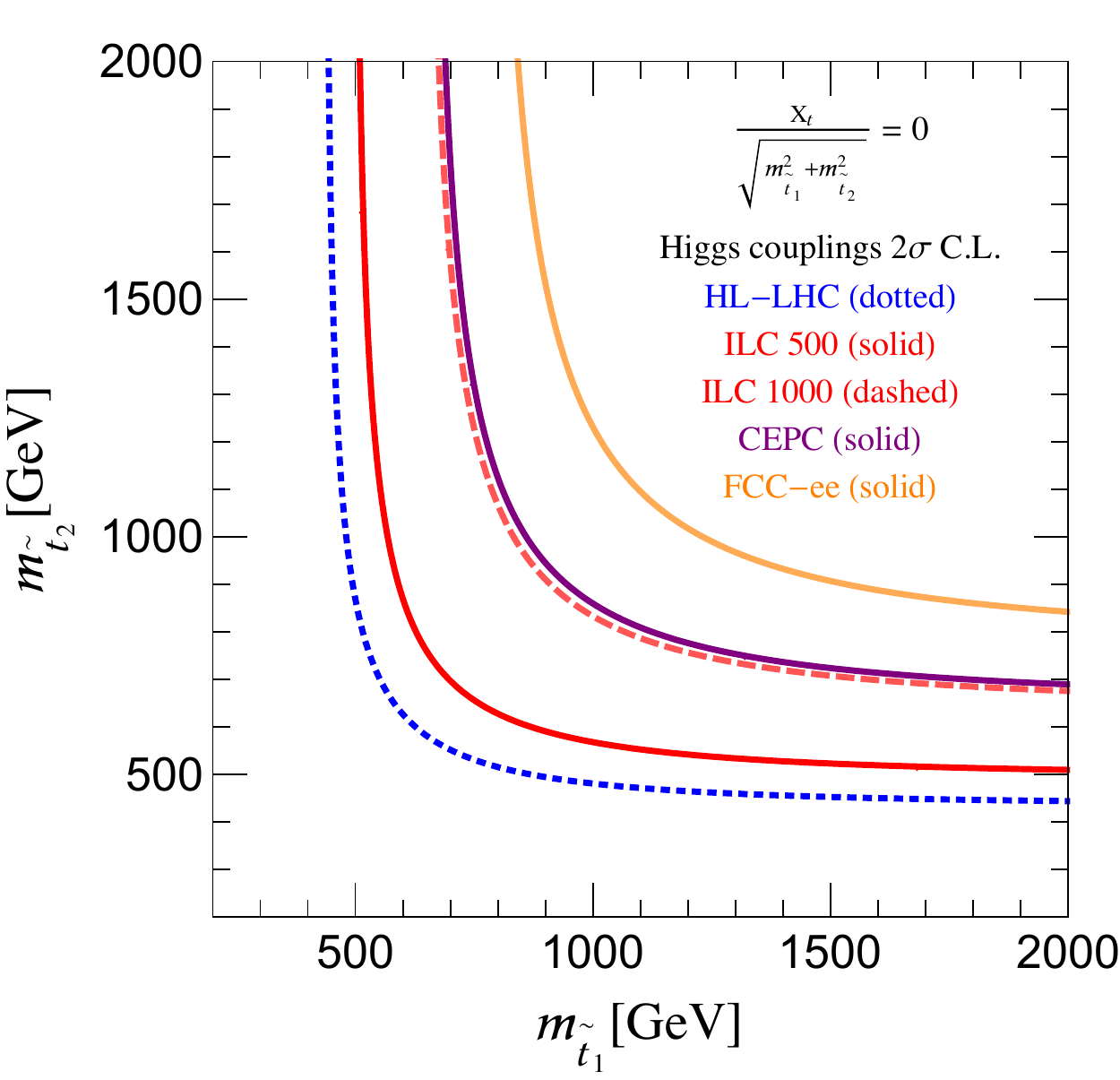} \\
\includegraphics[width=0.3\textwidth]{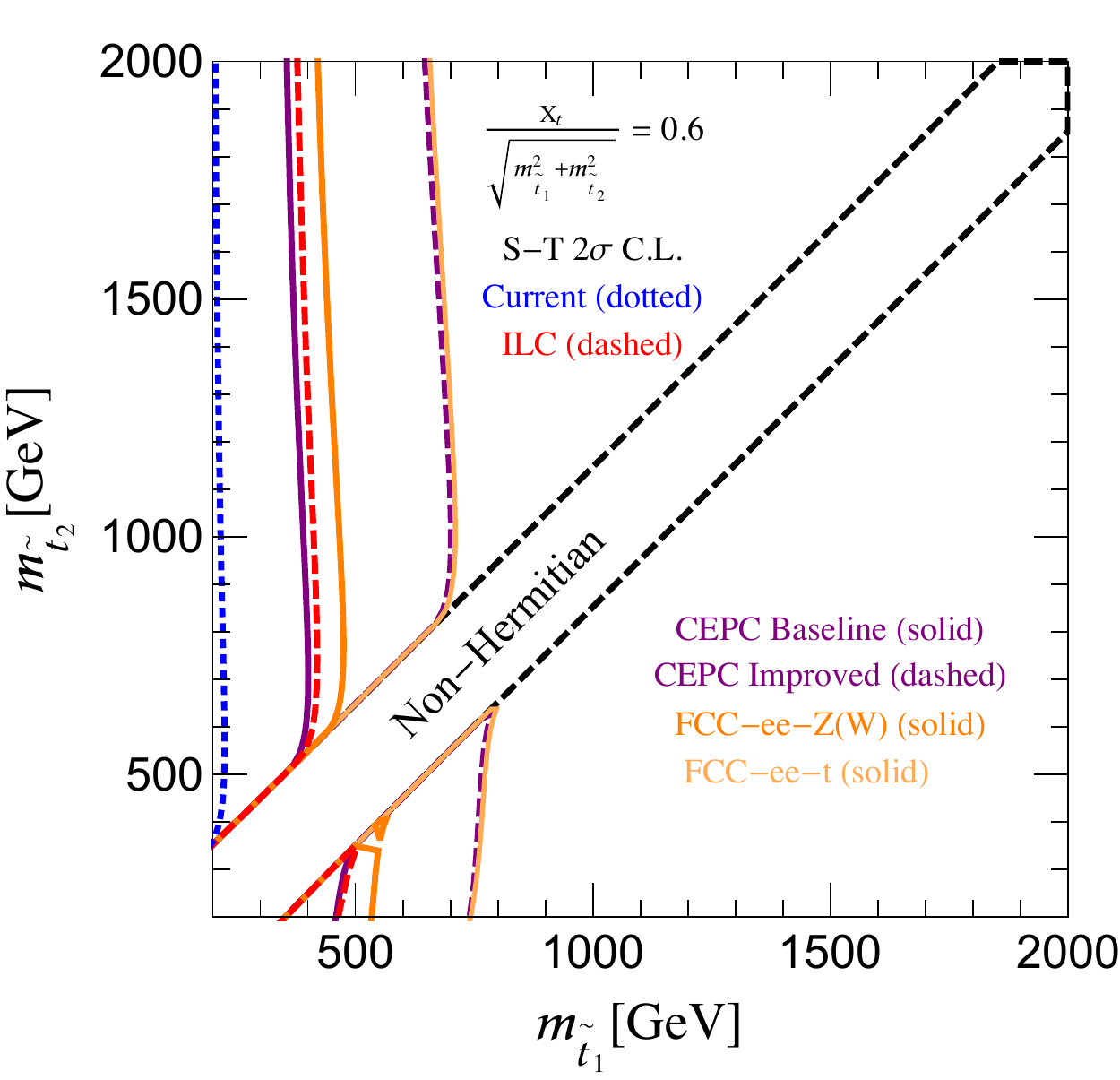}\quad  \includegraphics[width=0.3\textwidth]{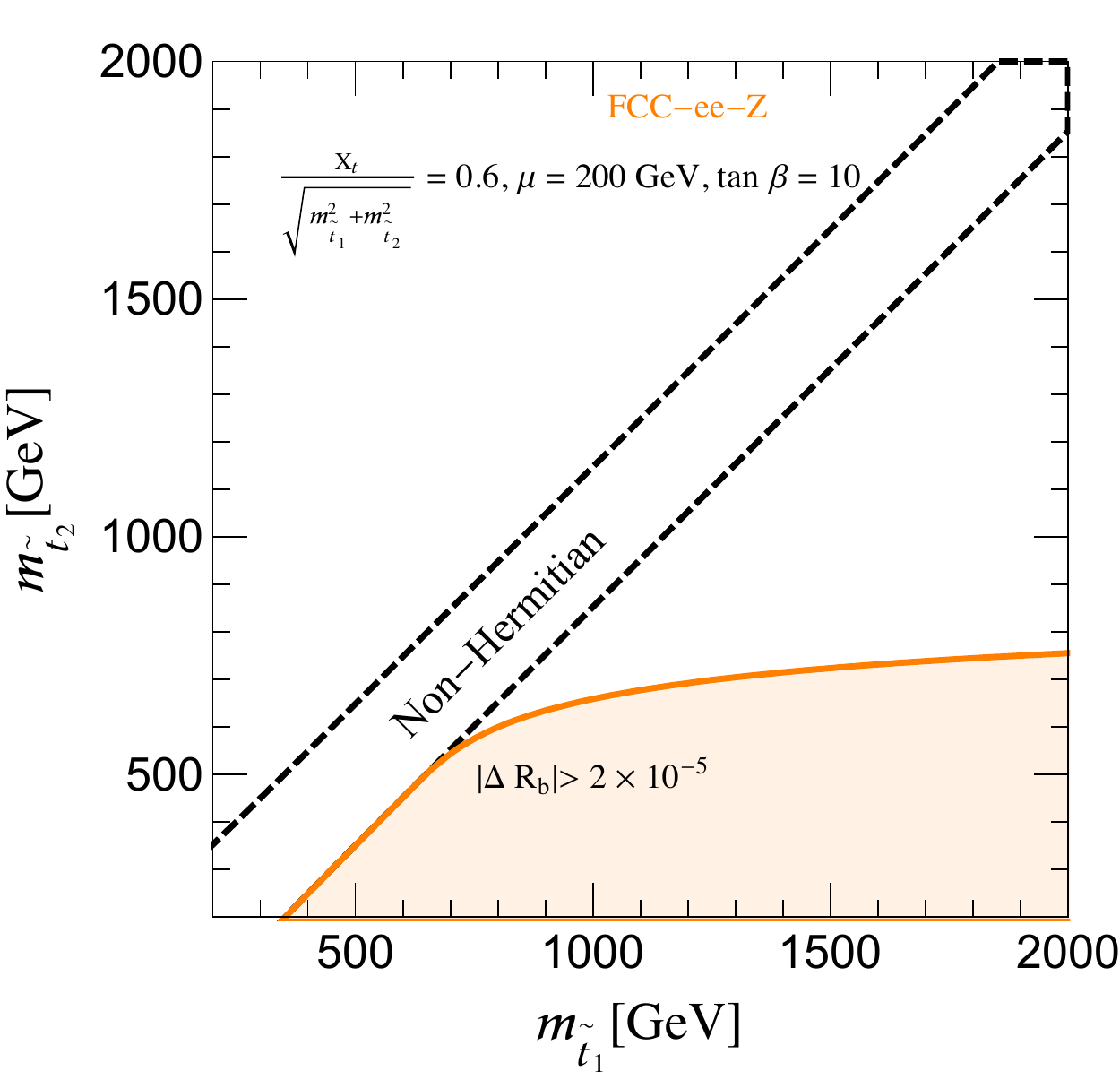} \quad  \includegraphics[width=0.3\textwidth]{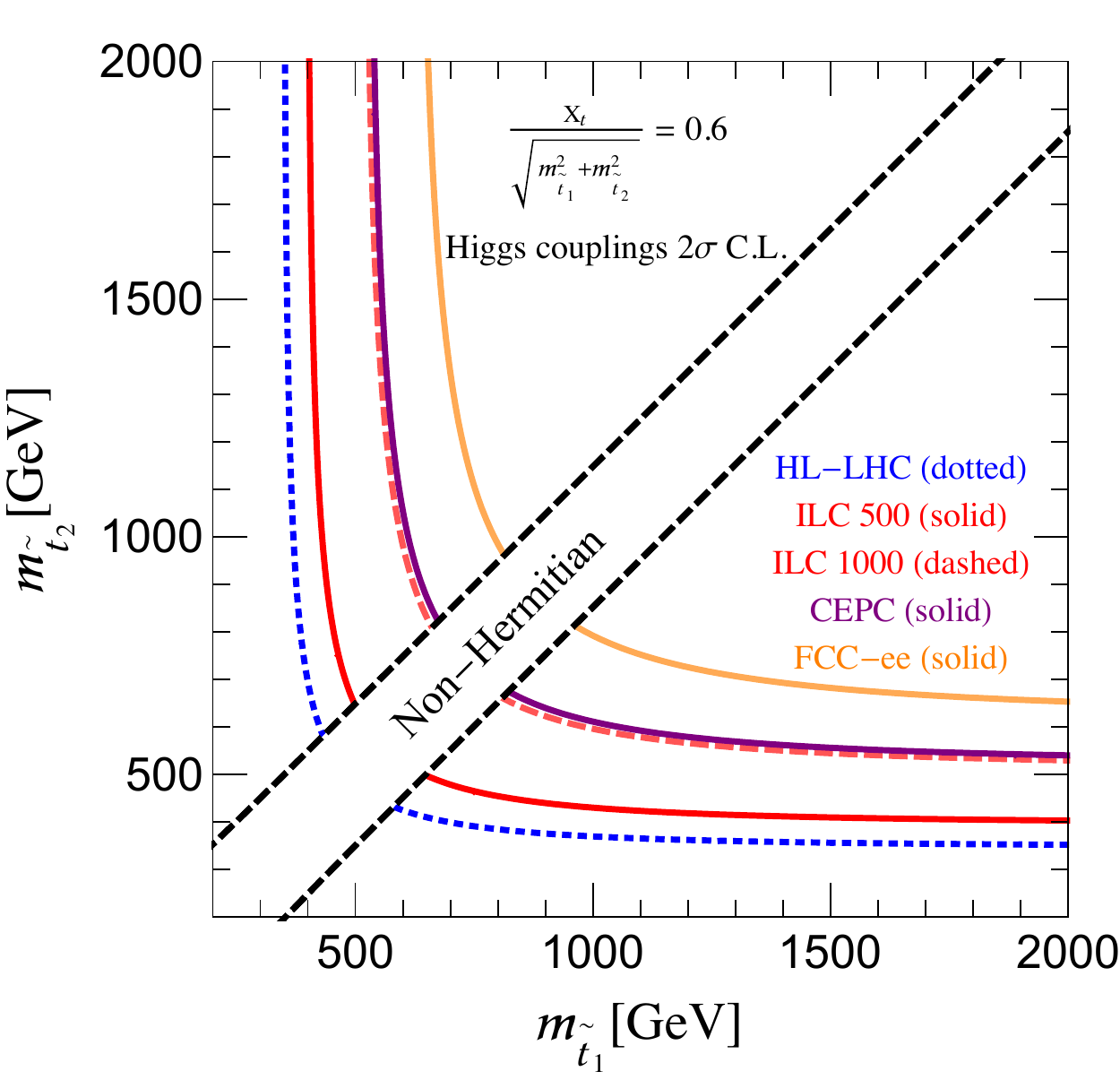}   \\
\includegraphics[width=0.3\textwidth]{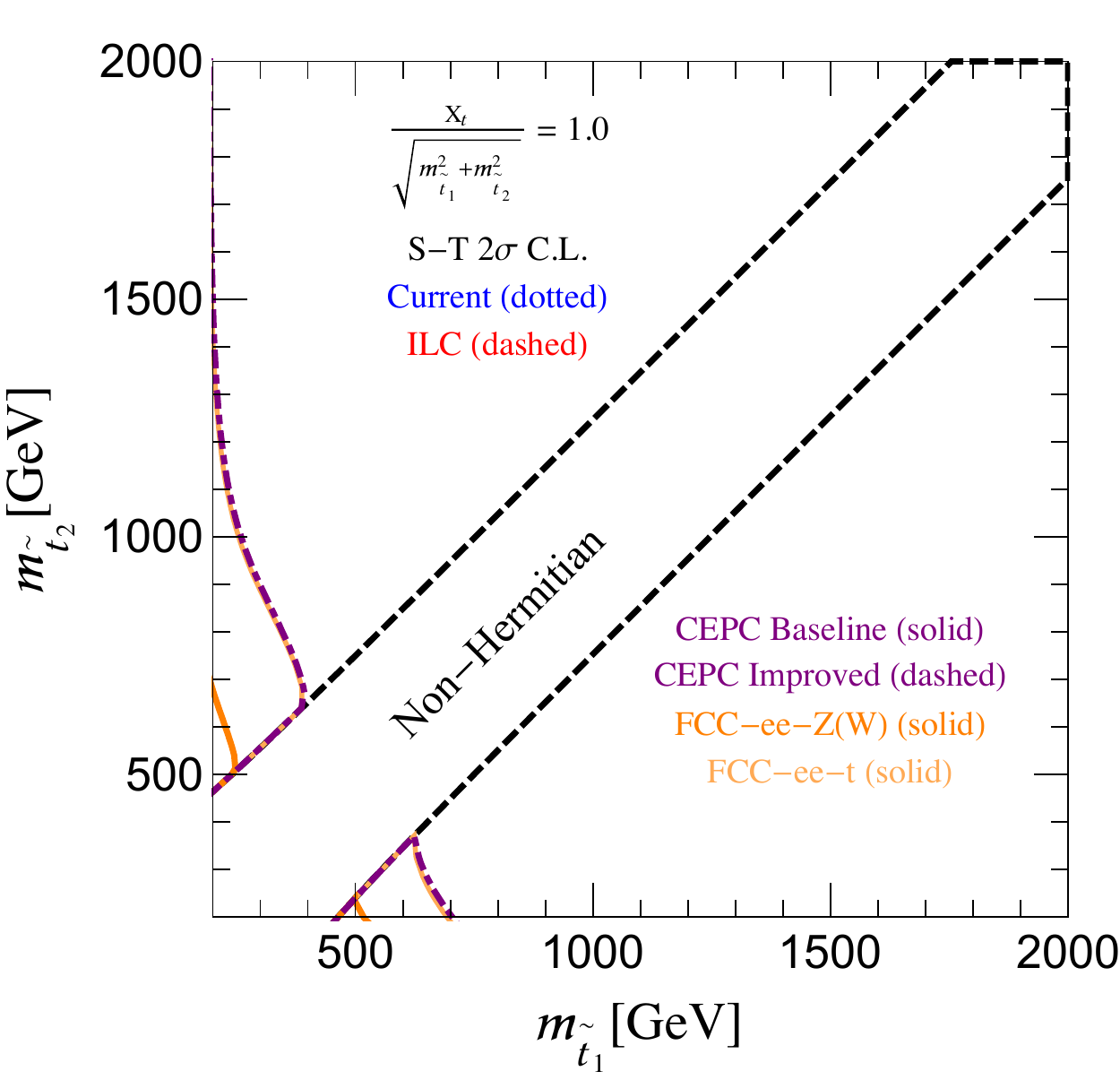}\quad  \includegraphics[width=0.3\textwidth]{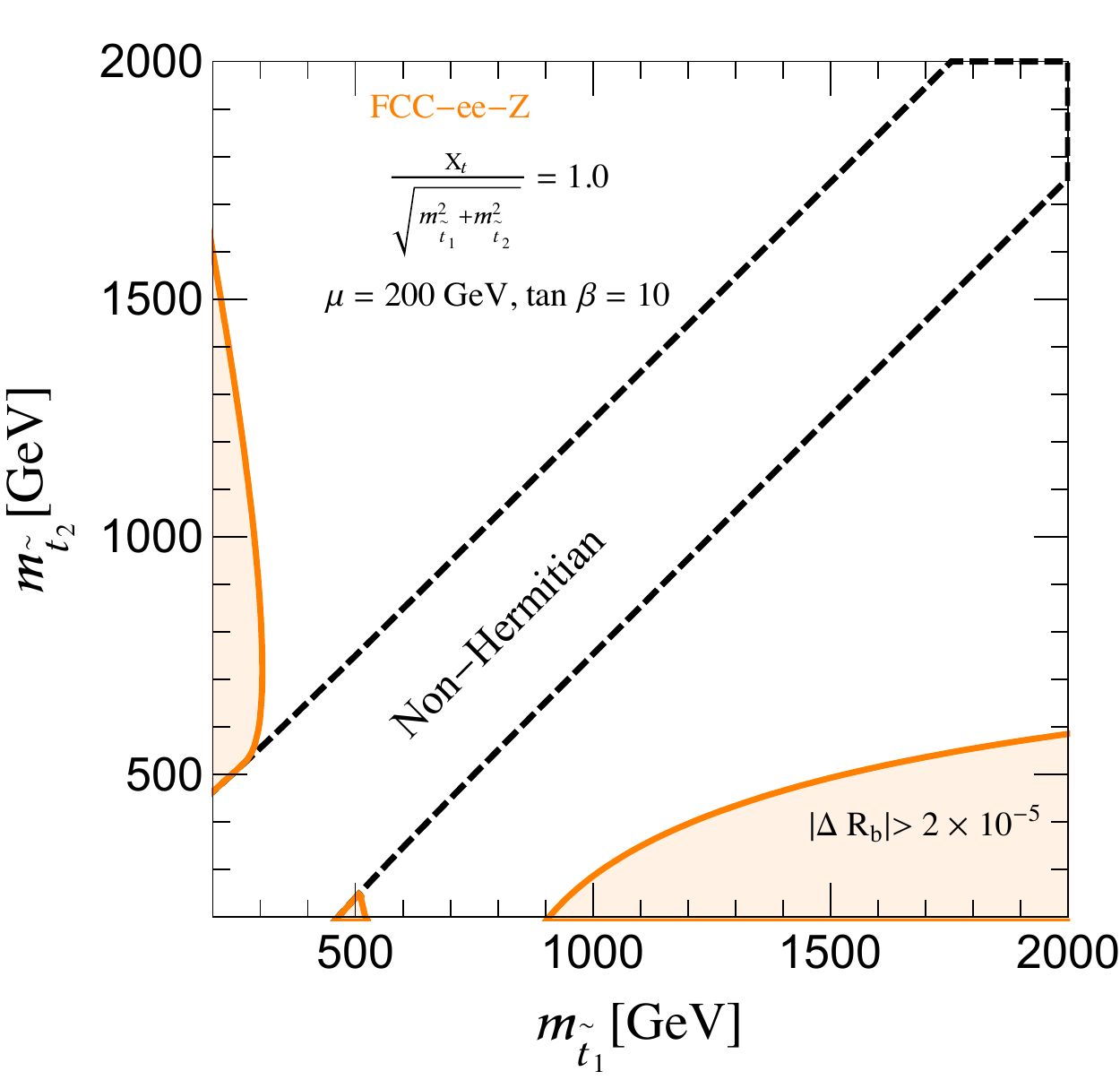} \quad  \includegraphics[width=0.3\textwidth]{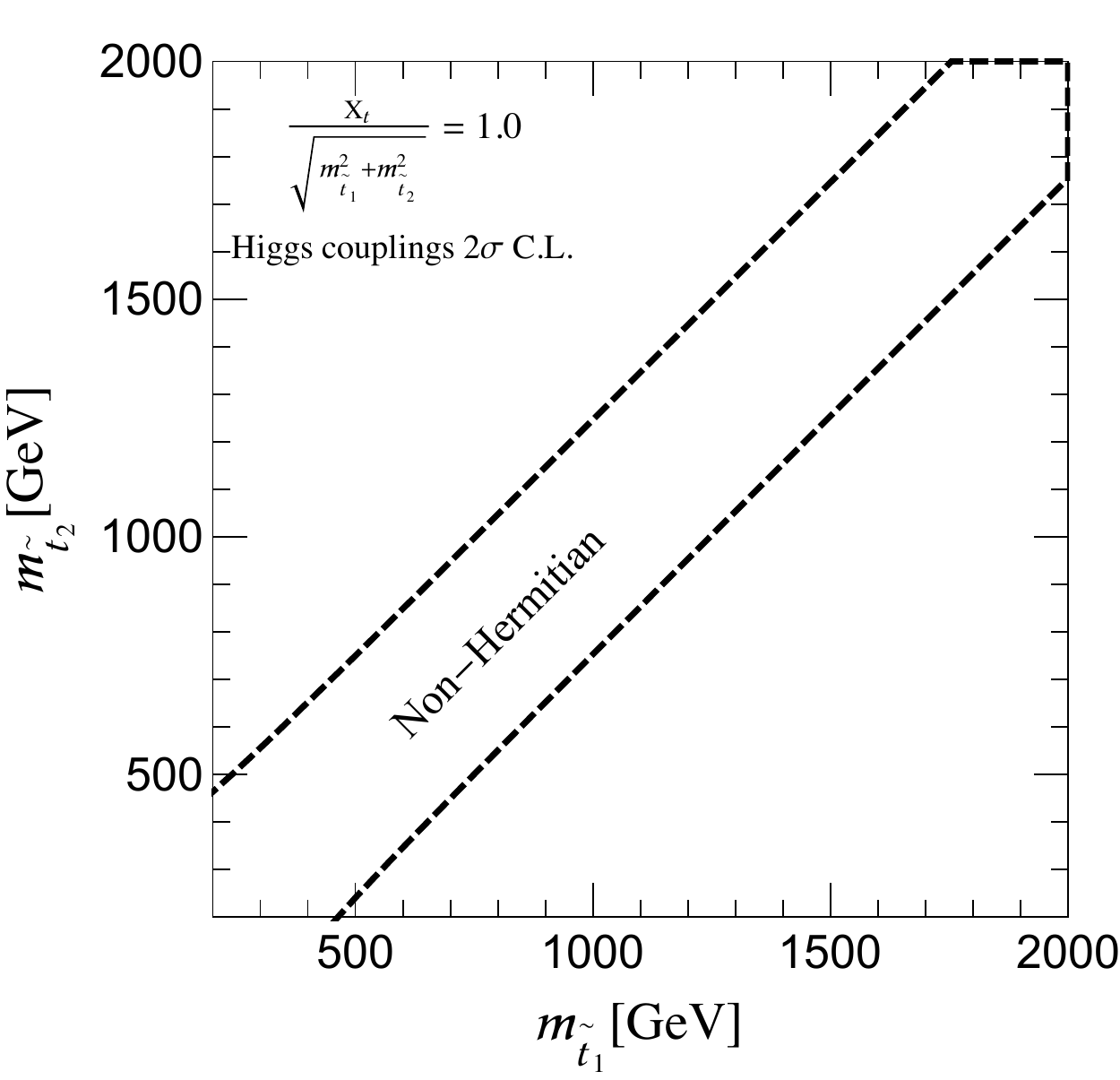}  \\
\includegraphics[width=0.3\textwidth]{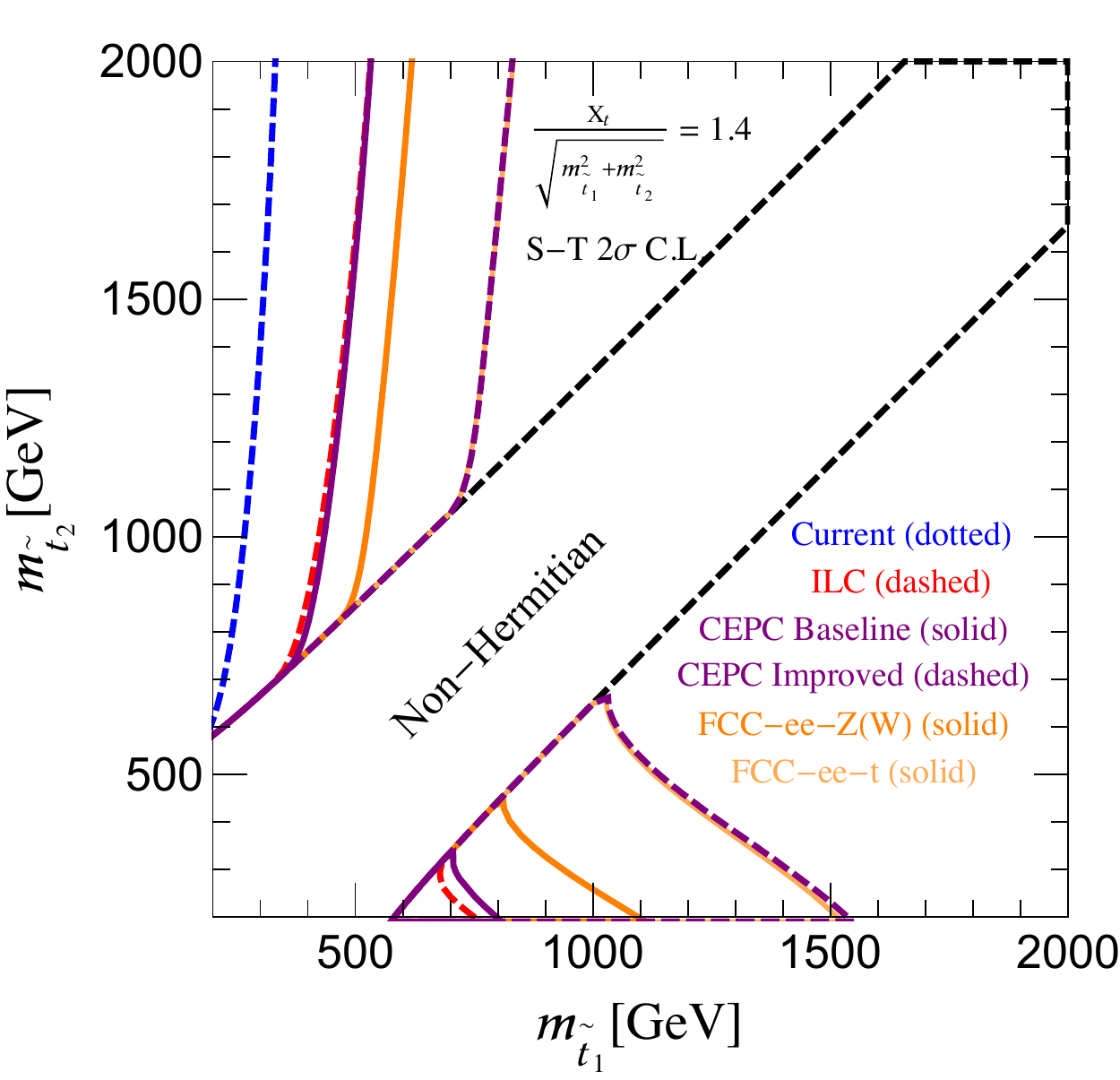}\quad  \includegraphics[width=0.3\textwidth]{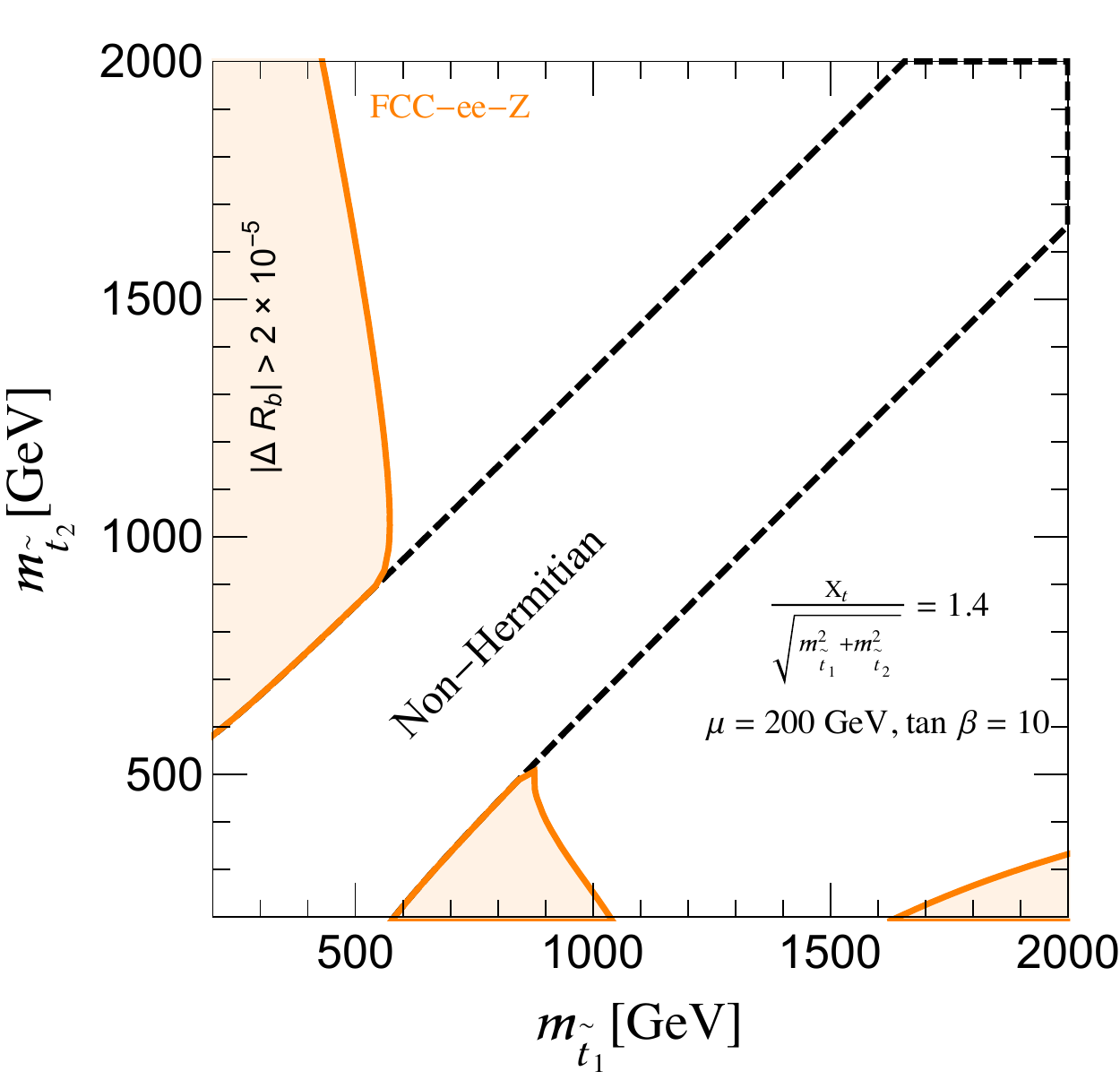} \quad  \includegraphics[width=0.3\textwidth]{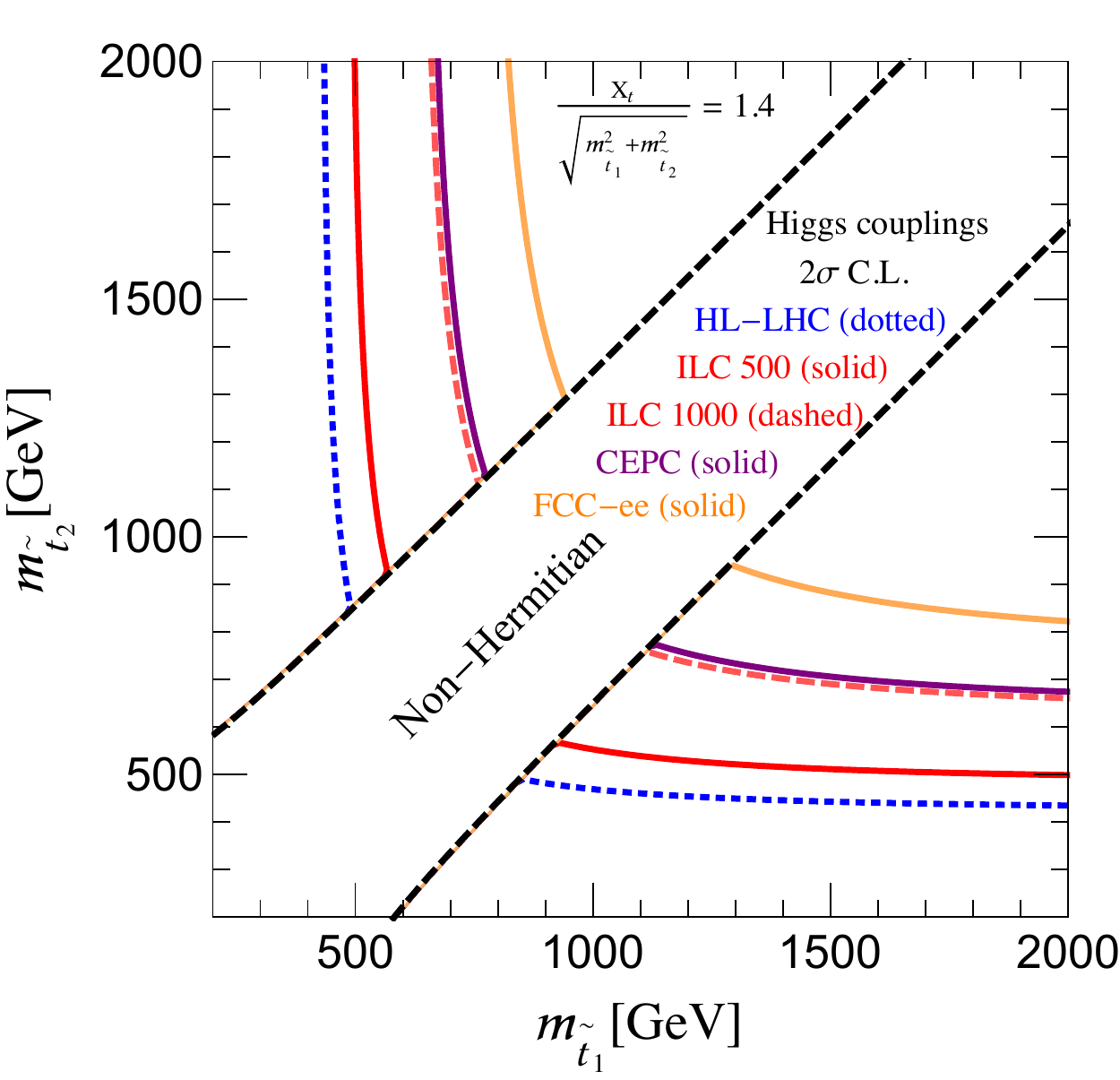}
\end{center}
\caption{Regions in the stop physical mass plane that are/will be excluded at 2$\sigma$ by EWPT with oblique corrections (left column), $R_b$ at FCC-ee (mid column) and Higgs couplings (right column) for different choices of $X_t/\sqrt{m_{\tilde{t}_1}^2+m_{\tilde{t}_2}^2}$: 0 (first row), 0.6 (2nd row), 1.0 (3rd row) and 1.4 (last row). We chose the mass eigenstate with $m_{\tilde{t}_1}$ to be mostly left-handed while the mass eigenstate with $m_{\tilde{t}_2}$ to be mostly right-handed. For non-zero choices of $X_t$, there are regions along the diagonal line which cannot be attained by diagonalizing a Hermitian mass matrix~\cite{Fan:2014txa}. Also notice that the vacuum instability bound constrains $X_t/\sqrt{m_{\tilde{t}_1}^2+m_{\tilde{t}_2}^2} \lesssim \sqrt{3}$~\cite{Blinov:2013fta}. }
\label{fig:result}
\end{figure}%
%%%%%%%%%%%%%%%%%%%%%%%%%%%%%%%%%%%%%%%%%%%%%%%%%%%%%

Now we turn to study the sensitivities of future EWPT to the stop sector. As we discussed in Sec.~\ref{sec:loopsST}, the mass splitting between left-handed stops and sbottoms violates custodial SU(2) and will generate a correction to the $T$ parameter, whereas the $S$ parameter correction is relatively small~\cite{Drees:1991zk}. Detailed formulas are in Appendix~\ref{app:loops}. The mixing between left- and right-handed stops would introduce some cancelation between various terms in Eq.~\ref{eq:Tstop}. It has been demonstrated in Ref.~\cite{Espinosa:2012in} that when the ratio between either the two stop or stop/sbottom masses is smaller than 3, the stop contribution to the $T$ parameter is minimized at about
\beq
\sin(2 \theta_{\tilde{t}}) \approx \frac{2m_t}{m_{\tilde{t}_2}-m_{\tilde{t}_1}}, \quad {\rm or~equivalently,} \quad X_t \approx m_{\tilde{t}_2}+m_{\tilde{t}_1}.  
\label{eq:minimizeTcorrection}
\eeq

The $S-T$ constraints from the analysis in the previous section could be translated into constraints on the parameters in the stop sector, which are shown in the left column of Fig.~\ref{fig:result}. When there is no mixing between stops, i.e. $X_t = 0$, current data already rules out the left-handed stop up to about 350 GeV. The ILC/GigaZ, CEPC baseline or FCC-ee/TeraZ program could push the limit on the left-handed stop up to about 600 GeV. With a top threshold scan at FCC-ee or an improved CEPC plan, the bound will be pushed to above 1 TeV. Once stop mixing is turned on, the bound will be relaxed. In particular, close to $X_t \approx m_{\tilde{t}_2}+m_{\tilde{t}_1}$, the bound on the stop masses almost vanishes as demonstrated in the third plot of the left column. At large mixing, when $X_t$ is above the sum of two physical masses, the constraints will reappear. Instead of containing mostly left-handed stops, the constraints begin to symmetrize for left- and right- handed stops. Notice that the vacuum instability bound constrains $X_t/\sqrt{m_{\tilde{t}_1}^2+m_{\tilde{t}_2}^2} \lesssim \sqrt{3}$~\cite{Blinov:2013fta} (the early analysis of the vauum instability bound constrains $X_t/\sqrt{m_{\tilde{t}_1}^2+m_{\tilde{t}_2}^2} \lesssim \sqrt{7}$~\cite{Kusenko:1996jn}). 

%%%%%%%%%%%%%%%%%%%%%%%%%%%%%%%%%%
\section{Non-oblique Corrections from Stops to $R_b$ }
\label{sec:Rb}
%%%%%%%%%%%%%%%%%%%%%%%%%%%%%%%%%%
\afterpage{\clearpage}

As discussed in Sec.~\ref{subsec:Rbops}, in addition to modifying the SM gauge boson two-point functions right-handed stops could also generate a non-oblique correction to the three-point function $Zb\bar{b}$ and thus the ratio of the partial width of $Z$ decaying to bottom quarks to the $Z$ hadronic partial width, denoted by $R_b$. Currently $R_b$ is measured to be $0.21629\pm 0.00066$ at LEP and SLC~\cite{ALEPH:2005ab}. The error bar is roughly equally shared by the systematic and statistical uncertainties. At ILC, the GigaZ running will accumulate $10^{10}$ $Z$'s in three years, which is a factor of $10^3$ times the statistics accumulated at LEP, and thus will reduce the statistical error by a factor of $30$. This makes the statistical uncertainty negligible compared to the more important systematic uncertainties. These include $b$-tagging capabilities, which are expected to be improved at ILC. At LEP, the individual experiments that have the smallest systematics uncertainty of $R_b$ measurements are DELPHI with $b$ tagging efficiency of $30\%$ and SLD with $b$ tagging efficiency of $62\%$ for almost pure $b$-jets~\cite{ALEPH:2005ab}. At ILC, the efficiency could be raised up to 80\%~\cite{Behnke:2001qq} and the precision of $R_b$ is expected to be improved by a factor of five relative to the result of LEP~\cite{Hawkings:1999ac,AguilarSaavedra:2001rg}, giving an uncertainty of about $1.3\times10^{-4}$. The CEPC $R_b$ measurement expects a 10 - 15\% higher $b$ tagging efficiency compared to the LEP one~\cite{LiangTalk}. This leads to a total uncertainty in $R_b$ of about $1.7\times10^{-4}$, which is dominated by the systematic error~\cite{LiangTalk}.

At FCC-ee, the statistics uncertainty will be reduced by a factor of 300 and the $b$-tagging capabilities are also expected to be similar to or better than those of SLD with a more granular vertex detector and a smaller beam spot. It's projected that FCC-ee could measure $R_b$ with a precision $(2 - 5) \times 10^{-5}$~\cite{Gomez-Ceballos:2013zzn}. It's more precise compared to the ILC and CEPC measurements though it should be understood that these numbers, especially those for CEPC and FCC-ee, may not be the final ones. Below we present an estimate of the reach for the stops using the most optimistic number.
 The current theory uncertainty of $R_b$ is $1.5 \times 10^{-4}$ from two-loop diagrams without closed fermion loops and higher-order contributions~\cite{Freitas:2014hra}. Completing the two-loop and the three-loop computations will bring the theory uncertainty down to a few times $10^{-5}$. Thus to achieve the precision estimate of FCC-ee, higher-order calculations of the total and partial widths of $Z$ are crucial. 

We will focus on the stop--charged higgsino loop as the stops and higgsinos are the necessary ingredients of natural SUSY. Even if contributions from other superparticles could cancel the stop--higgsino contribution to $R_b$, they will bring in additional fine-tunings. 
The full loop formulas for the stop--higgsino contribution to $R_b$ are presented in Appendix~\ref{app:Rb}, along with an approximate expansion when $\mu \ll m_{{\tilde t}_1}, m_{{\tilde t}_2}$. In the middle column of Fig.~\ref{fig:result}, we demonstrate regions in the physical stop mass plane with $|\Delta R_b| > 2 \times 10^{-5}$ fixing the Higgsino mass to be 200 GeV. Without mixing, measurement of $R_b$ with a precision of $2 \times 10^{-5}$ at FCC-ee could probe right-handed stops up to about 800 GeV. Similar to the $T$ parameter, the sensitivity decreases once the mixing is turned on and almost vanishes around $X_t = \sqrt{m_{\tilde{t}_1}^2+m_{\tilde{t}_2}^2}$. It increases again when mixing gets larger but constrains both left- and right-handed stops at the same time. 

%%%%%%%%%%%%%%%%%%%%%%%%%%%%%%%%%%%%%%%%%%%%%%%%%%%%%%
 \begin{figure}[!h]\begin{center}
\includegraphics[width=0.45\textwidth]{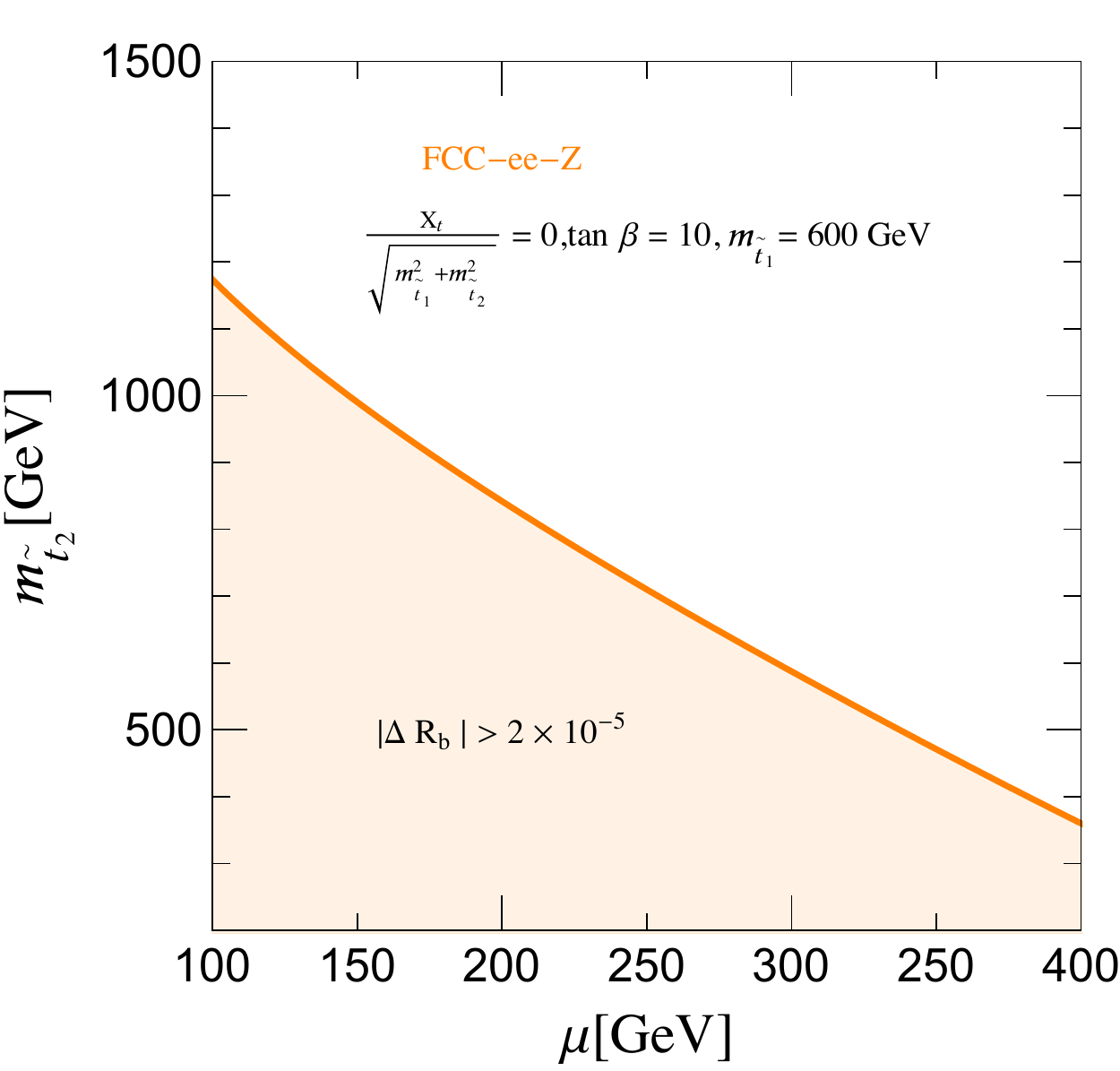} 
\end{center}
\caption{Region in the right-handed stop and Higgsino mass plane that the FCC-ee $R_b$ measurement could be sensitive to. We fix stop mixing to be zero, $\tan \beta = 10$ and left-handed stop mass to be 600 GeV. The result remains unchanged if one varies the left-handed stop mass.}
\label{fig:Rb}
\end{figure}%
%%%%%%%%%%%%%%%%%%%%%%%%%%%%%%%%%%%%%%%%%%%%%%%%%%%%%

Given the current, ILC and CEPC measurements of $R_b$, there are no constraint on stops. Only the most optimistic FCC-ee planned reach could achieve a significant constraint on right-handed stops. If $R_b$'s precision is $5 \times 10^{-5}$ instead of $2 \times 10^{-5}$, the constraint will be reduced to 350 GeV for zero stop mixing and Higgsino mass 200 GeV. This shows the importance of achieving a small beam spot and good $b$-tagging capabilities so that systematic uncertainties can be reduced as much as possible.

We also plot the constraint of the right-handed stop mass as a function of Higgsino mass $\mu$ in Fig.~\ref{fig:Rb}. One could see from the figure that the smaller $\mu$ is, the more sensitive $R_b$ measurement is to the right-handed stop. 

%%%%%%%%%%%%%%%%%%%%%%%%%%%%%%%%%%
\section{Higgs Coupling Constraints on Stops }
\label{sec:higgscoupling}
%%%%%%%%%%%%%%%%%%%%%%%%%%%%%%%%%%
Stop loops modify the Higgs couplings to gluons and photons, as we reviewed in Sec.~\ref{sec:higgsgammaglue}. Thus a precise measurement of Higgs digluon and diphoton couplings will indicate the degree of naturalness associated with stops. It is demonstrated in~\cite{Fan:2014txa} that independent of the stop mixing, current Higgs coupling data has excluded scenarios with both stops lighter than 400 GeV in the absence of Higgs mixing effects and suggests a minimum electroweak fine-tuning of between a factor of 5 and 10. 

In the right column of Fig.~\ref{fig:result}, we plot the regions that current and future Higgs coupling measurements could exclude at 2$\sigma$ C.L. We performed a profile likelihood {\bf one-parameter} fit to the estimated precisions of cross section and cross section times branching ratio for each future Higgs program, in which we only allow the Higgs--gluon coupling to vary and Higgs--photon coupling to vary in a correlated way dictated by Eq.~(\ref{eq:gamG}). Then we obtain the allowed range of $r_G^{\tilde{t}}$ at 2$\sigma$ C.L, which we map onto the stop parameter plane. The ILC precisions are tabulated in~\cite{Asner:2013psa}, CEPC precisions are estimated in~\cite{Fan} and FCC-ee ones could be found in the Snowmass Higgs working group report~\cite{Dawson:2013bba}. Notice that one {\bf should not} use the results of $\kappa_g$ from the seven-parameter fits which allow all Higgs couplings to vary freely~\cite{Dawson:2013bba}, as this will underestimate the exclusion. In the particular scenario we are considering, the variations of the Higgs couplings are much more constrained. For the ILC, we used the numbers of the ILC 500 scenario with the machine running at 250 GeV and 500 GeV with luminosities of 1150 fb$^{-1}$ and 1600 fb$^{-1}$ and the 1000 scenario with the machine running at 1 TeV in addition to the 500 case with a luminosity of 2500 fb$^{-1}$. For FCC-ee, the number assumes the machine running at 240 GeV and 350 GeV with luminosities of $10^4$ fb$^{-1}$ and 2600 fb$^{-1}$. From Fig.~\ref{fig:result}, one could see that the FCC-ee scenario is the most sensitive case. Again at the special point $X_t \sim \sqrt{m_{\tilde{t}_1}^2+ m_{\tilde{t}_2}^2}$, $r_G^{\tilde t} \approx 0$ from Eq.~\ref{eq:rG} and the bound vanishes. 

The strongest limit on the stop parameters comes from the measurement of $hgg$ coupling. This is due to a combination of the large size of the correction and the high precision of the measurements of this coupling at the Higgs factories.

%%%%%%%%%%%%%%%%%%%%%%%%%%%%%%%%%%
\section{The Light Stop Blind Spot}
\label{sec:blindspot}
%%%%%%%%%%%%%%%%%%%%%%%%%%%%%%%%%%

\afterpage{\clearpage}

It is apparent from Fig.~\ref{fig:result} that in the case $X_t^2 \approx m_{{\tilde t}_1}^2 + m_{{\tilde t}_2}^2$, all of the precision loop observables we consider have a significantly poorer reach than for other choices of $X_t$. This is a ``blind spot'' for precision tests of light stops. In calling this choice of $X_t$ a blind spot, we follow the terminology of ref.~\cite{Cheung:2012qy}, which coined the term for regions of neutralino parameter space that evade direct detection experiments. The analogy is a close one: the neutralino blind spots exist when the lightest neutralino has a vanishing tree-level coupling to the Higgs boson. The underlying reason for the blind spot in stop detection is that the {\em lightest} stop mass eigenstate has a vanishing tree-level coupling to the Higgs boson. In this case, the heavy stop can still contribute to precision observables, but its contributions are relatively small due to the larger mass suppression. (While this draft was being finalized, the blind spot region of parameter space was independently pointed out in ref.~\cite{Craig:2014una}. The fact that stops' contribution to $hgg$ coupling vanishes in the blind spot is well known and has been studied before, for example, in~\cite{Guo:2013iij}.)

To understand where the blind spot occurs, we can integrate out the heavy stop mass eigenstate ${\tilde t}_h$ to determine an effective quartic coupling of the light stop ${\tilde t}_l$ to the Higgs boson:
\beq
\begin{tikzpicture}[line width=1.5 pt]
\draw[scalar] (0,0)--(1.5,0);
\draw[scalar] (1.5,0)--(3,0);
\draw[scalar] (1.0,2.0)--(1.5,0);
\draw[scalar] (1.5,0)--(2.0,2.0);
\node at (3.5,1) {$+$};
\draw[scalar] (4,0)--(5.25,0);
\draw[scalar] (5.25,0)--(6.75,0);
\draw[scalar] (6.75,0)--(8,0);
\draw[scalar] (4,2)--(5.25,0);
\draw[scalar] (6.75,0)--(8,2);
\node at (-0.1,-0.25) {${\tilde t}_l$};
\node at (3.1,-0.25) {${\tilde t}_l$};
\node at (0.5,2.0) {$h$};
\node at (2.5,2.0) {$h$};
\node at (1.5,-0.25) {$y_t^2$};
\node at (3.9,-0.25) {${\tilde t}_l$};
\node at (6,0.5) {${\tilde t}_h$};
\node at (8.1,-0.25) {${\tilde t}_l$};
\node at (4.5,2) {$h$};
\node at (7.5,2) {$h$};
\node at (5.25,-0.25) {$y_t X_t$};
\node at (6.75,-0.25) {$y_t X_t$};
\end{tikzpicture}
\eeq
This leads to an effective coupling:
\beq
{\cal L}_{\rm eff} =  \left(y_t^2 - \frac{y_t^2 X_t^2}{m_{{\tilde t}_h}^2 - m_{{\tilde t}_l}^2}\right) \left|H_u\right|^2 \left|{\tilde t}_l\right|^2. 
\eeq
This leads to the ``blind spot'' mixing for which the coupling of the light stop to the Higgs boson vanishes:
\beq
X_t^* = \left(m_{{\tilde t}_h}^2 - m_{{\tilde t}_l}^2\right)^{1/2}.
\label{eq:Xtblindspot}
\eeq
This is also apparent from Eq.~\ref{eq:stopstophiggs}. Alternatively, one could find this critical mixing by evaluating the light stop mass eigenvalue and solving  the equation $\partial \log m_{{\tilde t}_l}/\partial \log v = 0$ for $X_t$.

We noted in Eq.~\ref{eq:minimizeTcorrection} that the $T$ parameter correction is minimized when $X_t$ is approximately the sum of the two stop mass eigenvalues, whereas Eq.~\ref{eq:hgg} makes it clear that the $hgg$ and $h\gamma\gamma$ corrections are minimized when $X_t$ is approximately equal to the two stop mass eigenvalues added in quadrature. These results agree with Eq.~\ref{eq:Xtblindspot} to the extent that one stop is significantly heavier than the other, because they all reduce to $X_t \approx m_{{\tilde t}_h}$. On the other hand, if the two stop mass eigenvalues are very close together, $X_t$ is necessarily small and the constraints are unaffected by the mixing.

Although being in this blind spot may help hide the light stop at future lepton colliders, it will make the fine-tuning even worse. In the limit of vanishing coupling between the Higgs and the light stop, the fine-tuning is dominated by the mass of the heavy stop. Therefore, it is still significant, especially in the precise blind-spot limit, in which the heavy stop is much heavier than the light stop. Nevertheless, it is interesting to discuss whether the signal of light stop could be found in some other observables.

The approximate blind spot in $R_b$ arising for the same values of $X_t$ appears to be a numerical accident, because the loop diagram Fig.~\ref{fig:Zbbcorrection} does not involve the coupling of the light stop to the Higgs boson. Other precision observables that do not depend on the stop--Higgs coupling could potentially probe the blind spot, but as discussed in Sec.~\ref{sec:othercorrections}, most such operators have numerically small coefficients that are difficult to probe. Perhaps the best probe of the blind spot region is $b \to s \gamma$, which is strongly constraining for large $A_t  \tan \beta$~\cite{Barbieri:1993av,Okada:1993sx,Ishiwata:2011ab,Blum:2012ii,Espinosa:2012in,Altmannshofer:2012ks,Katz:2014mba}. A detailed discussion of this probe is presented in the next section.

%%%%%%%%%%%%%%%%%%%%%%%%%%%%%%%%%%
\section{Conclusions and Discussions}
\label{sec:discussions}
%%%%%%%%%%%%%%%%%%%%%%%%%%%%%%%%%%
\afterpage{\clearpage}

\subsection{Implications for Fine-Tuning}

\begin{figure}[t]\begin{center}
\includegraphics[width=\textwidth]{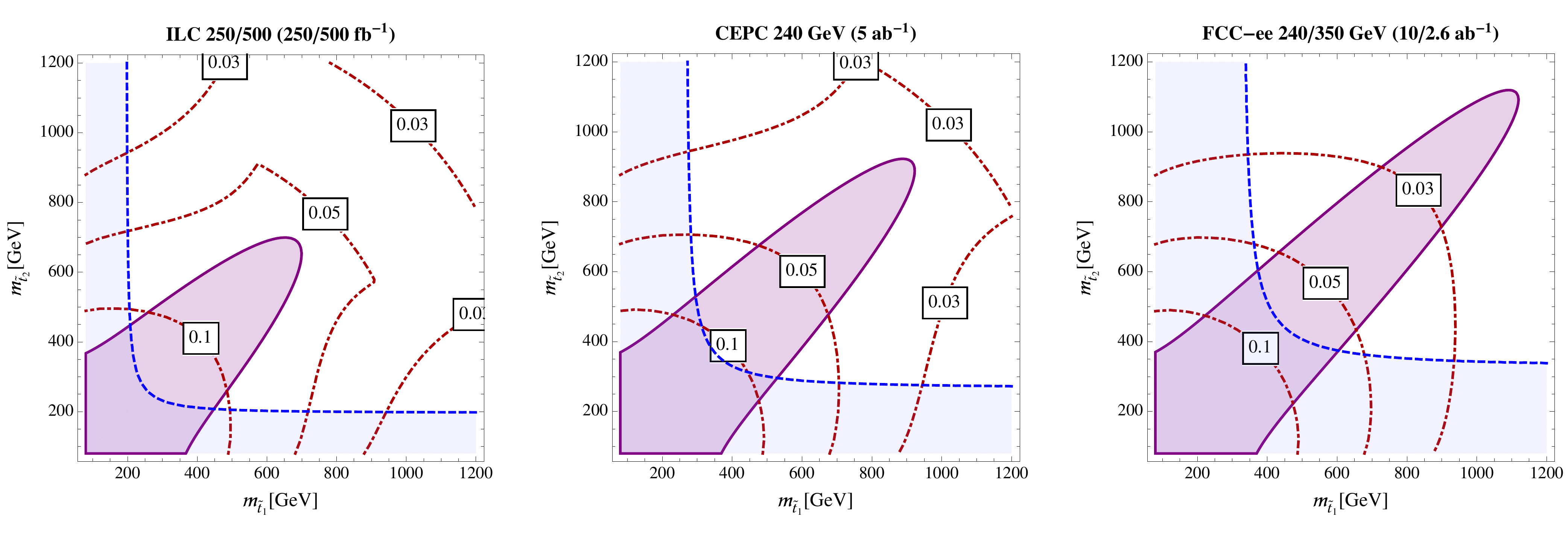}
\end{center}
\caption{Projected constraints in the stop mass plane from a one-parameter fit to the Higgs data from future experiments. The purple shaded region along the diagonal is excluded because the smallest $\left|X_t\right|$ consistent with the data at 2$\sigma$ is larger than the  maximum $\left|X_t\right|$ compatible with the mass eigenvalues, as explained in detail in ref.~\cite{Fan:2014txa}. The blue shaded region requires tuning $X_t$ to a part in 10 to fit the data. The dot-dashed red contours quantify fine-tuning in the Higgs mass from the quadratic sensitivity to stop soft terms.}
\label{fig:stopmassexclude}
\end{figure}

For a first look at fine-tuning, we show in Fig.~\ref{fig:stopmassexclude} a comparison of bounds in the plane of stop mass eigenvalues with contours of fixed fine-tuning in the Higgs mass arising from quadratic sensitivity to the stop soft masses. The Higgs mass fine-tuning from the stop sector is defined as~\cite{Kitano:2006gv,Perelstein:2007nx}
\beq\label{eq:Dz}
\left(\Delta_h^{-1}\right)_{\tilde t}=\left|\frac{2 \delta m_{H_u}^2}{m_h^2}\right|,\;\;\;\quad \delta m_{H_u}^2|_{\rm stop}&=&-\frac{3}{8\pi^2}y_t^2\left(m_{Q_3}^2+m_{U_3}^2+A_t^2\right)\log\left(\frac{\Lambda}{m_{\rm EW}}\right) \nonumber \\
&=&-\frac{3}{8\pi^2}y_t^2\left(m_{\tilde t_1}^2+m_{\tilde t_2}^2-2m_t^2+A_t^2\right)\log\left(\frac{\Lambda}{m_{\rm EW}}\right).
\eeq
Here $\Lambda$ is a scale characterizing mediation of SUSY breaking, while $m_{\rm EW}$ is the low-energy scale where the running stops. We take $m_{\rm EW} = \max(\sqrt{m_{{\tilde t}_1} m_{{\tilde t}_2}}, m_h)$. In Fig.~\ref{fig:stopmassexclude}, we take 
\beq
A_t = \max\left(0, \left|X_t^{\rm min} \right| + \mu/\tan \beta\right)
\eeq
with the SUSY breaking mediation scale $\Lambda = 30$ TeV, $\mu = 200$ GeV and $\tan\beta =10$. Here $\left|X_t^{\rm min}\right|$ is taken to be the smallest absolute value of $X_t$ allowed by the Higgs coupling measurements at 2$\sigma$ C.L. for given stop masses~\cite{Fan:2014txa}. We have chosen on purpose a very low SUSY mediation scale to draw conservative conclusions about the tuning level. This plot shows the region of parameter space which can be excluded by Higgs data {\bf independent of the stop mixing angle}, in purple, along with the region that is excluded only for a fine-tuned set of mixing angles, in blue. The blue curve corresponds a part-in-10 adjustment of $X_t$: $\left|m_{{\tilde t}_1}^2 + m_{{\tilde t}_2}^2 - X_t^2\right| < \frac{1}{10} \left|X_t\right|^2$. For a detailed discussion of how such a plot is computed, see Ref.~\cite{Fan:2014txa}. Fig.~7 of that reference inferred a bound from the constraints on $\kappa_g$ and $\kappa_\gamma$ listed in the seven-parameter fit of ref.~\cite{Dawson:2013bba}. We have improved the calculation by performing a one-parameter fit to all projected $\sigma$ and $\sigma \times {\rm Br}$ measurements, which slightly improves the reach. Specifically, the approach taken in Ref.~\cite{Fan:2014txa} was based on bounds that allowed other parameters to float, whereas here we extract stronger bounds by assuming that stops are the only contribution to the new physics. We also provide, for the first time, an estimate of the reach of CEPC. The combined ILC 250, 500, and 1000 GeV runs would have a very similar reach to CEPC.

%%%%%%%%%%%%%%%%%%%%%%%%%%%%%%%%%%%%%%%%%%%%%%%%%%%%%%
%%%%%%%%%%%%%%%%%%%%%%%%%%%%%%%%%%%%%%%%%%%%%%%%%%%%%

From this plot we see that any future Higgs factory would mostly or entirely rule out regions of 10\% fine tuning, but will leave gaps with 5\% fine tuning. These gaps occur due to the blind spot discussed above. As we have noted above, measurements of $b \to s\gamma$ can help to constrain the blind spot region. However, bounds from $b \to s\gamma$ depend not only on the stop mass matrix but also on $\mu$ and $\tan \beta$. To provide a perspective on the implications of these bounds for fine-tuning, we should assess the tree-level tuning arising from $\mu$ and from $m_A$.

The precise measurement of Higgs couplings to fermions is sensitive to the mass scale of the heavy Higgs bosons $A^0$, $H^0$, $H^\pm$ that are present in the MSSM and its extensions. Mixing among the Higgs bosons will always modify the coupling of the light Higgs to fermions at order $m_h^2 / m_A^2$. (We will collectively denote the masses of all of these particles as $m_A$, although there may be some splitting between $H^0$ and $A^0$.) The coefficient is somewhat model dependent. We can estimate the bound on these couplings by focusing on $\kappa_b$, which is well-measured and approximately equal to
\beq
\kappa_b  \equiv\frac{y_{hbb}^{{\rm SUSY}}}{y_{hbb}^{\rm SM}} \approx 1 + 2 \frac{m_h^2}{m_A^2}
\eeq
at large $\tan \beta$ in models where the dominant new quartic coupling beyond the MSSM arises from nondecoupling $D$-terms~\cite{Blum:2012ii,Gupta:2012fy,D'Agnolo:2012mj}. Models with new quartics arising from $F$-terms have a somewhat different structure, but would yield a similar bound on $m_A$ up to order-one factors (especially since $\tan \beta$ in theories like the NMSSM cannot be very large). Doing a one-parameter fit with only $\kappa_b$ deviating from one, we find the following $2\sigma$ bounds:
\beq
{\rm ILC-500}: \left|\kappa_b - 1\right| < 1.3\% & \Rightarrow & m_A > 1.5~{\rm TeV}, \\
{\rm CEPC}: \left|\kappa_b - 1\right| < 0.71\% & \Rightarrow & m_A > 2.1~{\rm TeV}, \\
{\rm FCC-ee}: \left|\kappa_b - 1\right| < 0.39\% & \Rightarrow & m_A > 2.8~{\rm TeV}.
\eeq
These bounds on $m_A$ imply moderate fine-tuning, unless $\tan \beta$ is large. We estimate the fine-tuning of the Higgs potential due to large $m_A$ to be~\cite{Perelstein:2007nx,Gherghetta:2014xea,Katz:2014mba}
\beq
\Delta_A \approx \frac{2 m_A^2}{m_h^2 \tan^2\beta}.
\eeq
This shows that a failure to observe a deviation in $\kappa_b$ will imply either moderate fine-tuning or moderately large values of $\tan \beta$. The other tree-level tuning arises from $\mu$~\cite{Barbieri:1987fn,Kitano:2006gv,Perelstein:2007nx}:
\beq
\Delta_\mu \approx \frac{4 \mu^2}{m_h^2}.
\eeq
The constraints from $b \to s\gamma$ depend on choices of $\mu$ and $\tan \beta$. They can be made weaker at small $\tan \beta$ at the cost of larger $\Delta_A$~\cite{Katz:2014mba}. They could also be made weaker by making $\mu$ large to suppress the loop function, but this increases $\Delta_\mu$. There is another possibility of large SUSY-breaking contributions to higgsino masses that do not affect the EWSB conditions, as from the operator $K \supset X^\dagger X D_\alpha H_u D_\alpha H_d$. For such an operator to be important, we would require very low-scale SUSY-breaking. This is an interesting possibility and one that may require more attention if it becomes the only unconstrained scenario without tuning.

%%%%%%%%%%%%%%%%%%%%%%%%%%%%%%%%%%%%%%%%%%%%%%%%%%%%%%
\begin{figure}[h]\begin{center}
\includegraphics[width=\textwidth]{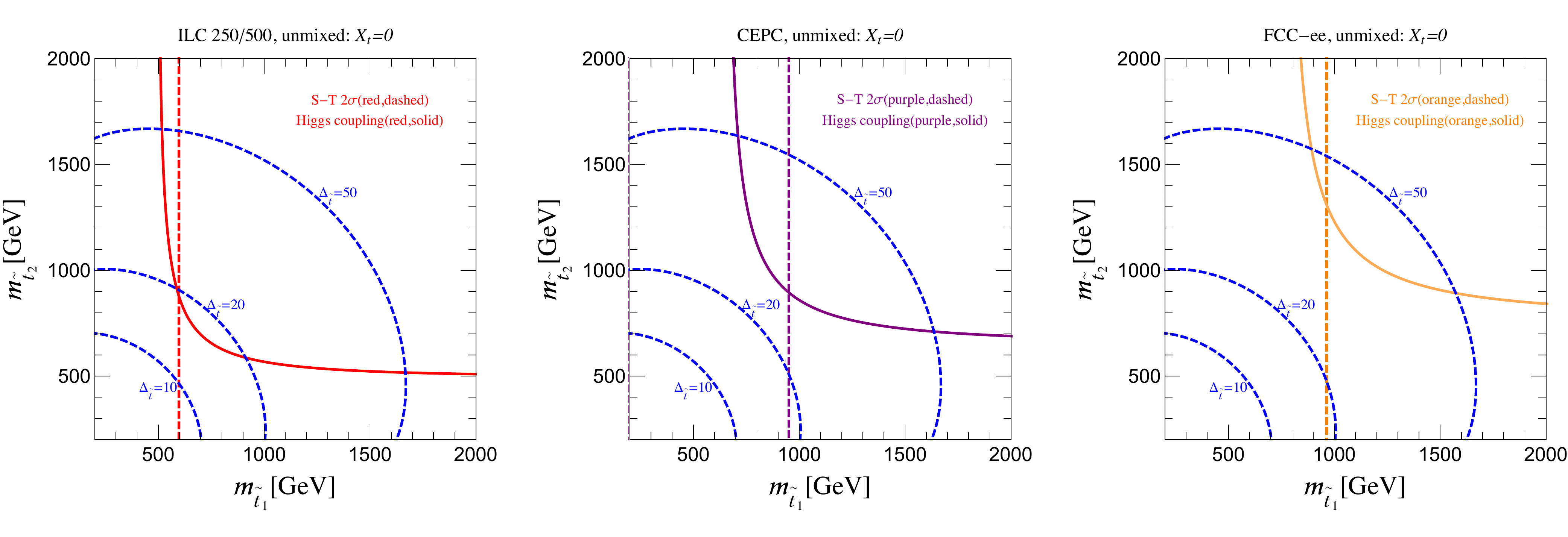}\\
\includegraphics[width=\textwidth]{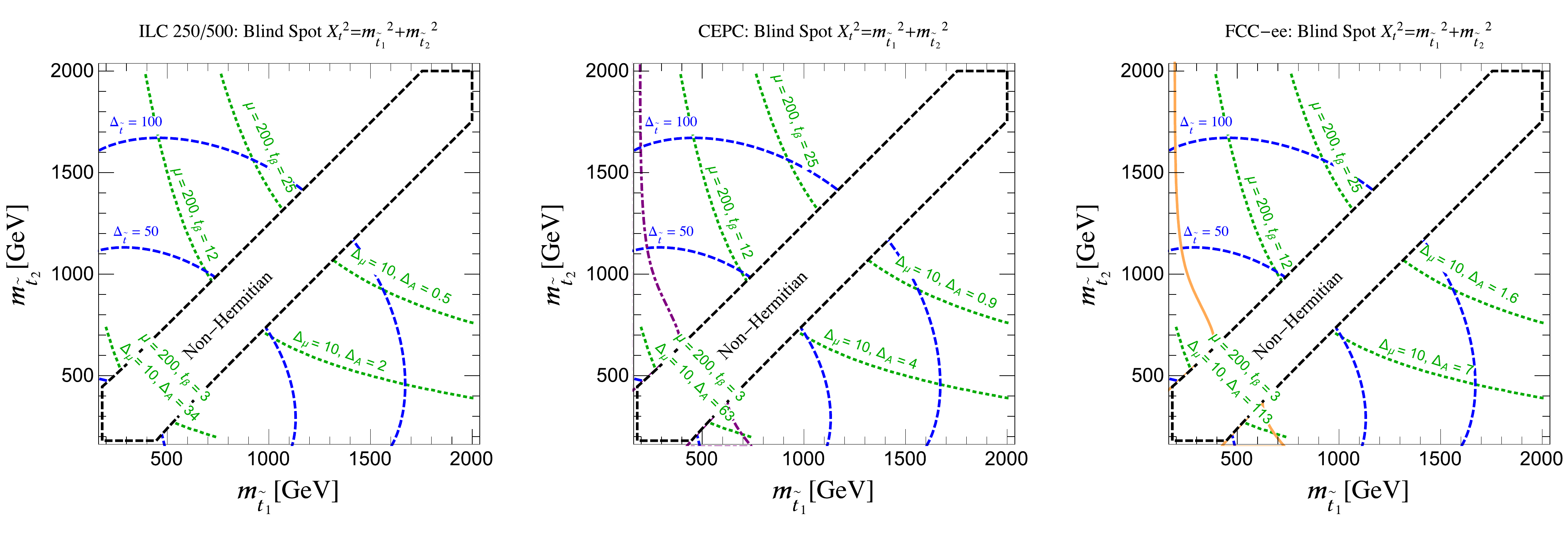}
\end{center}
\caption{Regions in the physical stop mass plane that precision measurements are sensitive to, with contours of tunings, at future $e^+e^-$ colliders (left: ILC; middle: CEPC; right: FCC-ee). Top row: bounds on stops with no mixing, $X_t=0$. Dashed vertical lines: 2$\sigma$ bounds on stop masses from $S$ and $T$ (mostly $T$); solid lines: 2$\sigma$ bounds on stop masses from Higgs coupling constraints. Blue dashed contours are the stop contributions to the Higgs mass tuning. Lower row: bounds on stops in the blind spot $X_t^2 = m_{{\tilde t}_1}^2 + m_{{\tilde t}_2}^2$. There are no Higgs measurement constraints. For CEPC with possible improvements  (purple dash-dotted line in the middle) or FCC-ee (orange solid line), EWPT is only sensitive to a small region. The green dashed lines are the exclusion contours from $b \to s\gamma$ for the choice $\mu = 200$ GeV and a few different values of $\tan \beta$. Each of these contours is also labeled with corresponding tunings $\Delta_\mu$ and $\Delta_A$. There is also a region along the diagonal line which cannot be attained by diagonalizing a Hermitian mass matrix~\cite{Fan:2014txa}.}
\label{fig:summaryplot}
\end{figure}%
%%%%%%%%%%%%%%%%%%%%%%%%%%%%%%%%%%%%%%%%%%%%%%%%%%%%%

Putting all of this together, we can summarize the implications of precision measurements for tuning in Fig.~\ref{fig:summaryplot}. The top row displays bounds on stops with no mixing ($X_t = 0$). We display the $2\sigma$ bounds on stop masses arising from EWPT (mostly the $T$-parameter) and from Higgs coupling constraints ($hgg$ and $h\gamma\gamma$), superimposed on contours of fixed stop contribution to the Higgs mass tuning. The fine-tunings are again computed using Eq.~\ref{eq:Dz} but with $X_t=0$ in this case.
From the figure we can see that the ILC would almost fully exclude regions with less than a factor of 20 tuning, whereas FCC-ee would reach almost to the factor of 50 tuning level. In the second row, we display constraints on the blind spot where $X_t^2 = m_{{\tilde t}_1}^2 + m_{{\tilde t}_2}^2$. In this case, the large $X_t$ will contribute more to the Higgs mass fine-tuning. One could see that from Eq.~\ref{eq:Dz} and by comparing the contours with the same Higgs mass tuning from stops in the first and second row of Fig.~\ref{fig:summaryplot}. 
Yet in this case Higgs coupling measurements are not constraining, and EWPT only exclude a small region at CEPC with possible improvements or at FCC-ee. However, $b \to s\gamma$ plays an interesting complementary role. We show exclusion contours (green dashed lines) from $b \to s\gamma$ for the choice $\mu = 200$ GeV and a few different values of $\tan \beta$. Each of these contours is also labeled with the corresponding tunings $\Delta_\mu$ and $\Delta_A$. From this we can see that the contour of low stop mass tuning ($\Delta_{\tilde t} = 10$), a blue dashed line which is barely visible at the lower left, is allowed only by going to $\tan \beta < 3$, at which point the tuning $\Delta_A$ will be large if no deviation has been observed in $\kappa_b$. If we restrict to large enough values of $\tan \beta$ to suppress $\Delta_A$, then the stop mass tuning $\Delta_{\tilde t}$ becomes large. In this way, the interplay between measurement of the Higgs couplings to fermions and the existing measurements of $b \to s\gamma$ will allow the blind spot region to be indirectly covered by future $e^+ e^-$ colliders. Notice that we deliberately choose a positive $\mu$ throughout the analysis. The sign of $\mu$ will only give a negligible modification to the calculation of the Higgs mass fine-tuning from the stops. However, for negative $\mu$, the $b\to s \gamma$ constraint will get considerably stronger.

\subsection{Implications for Folded Stops}

%%%%%%%%%%%%%%%%%%%%%%%%%%%%%%%%%%%%%%%%%%%%%%%%%%%%%%
\begin{figure}[h]\begin{center}
\includegraphics[width=0.66\textwidth]{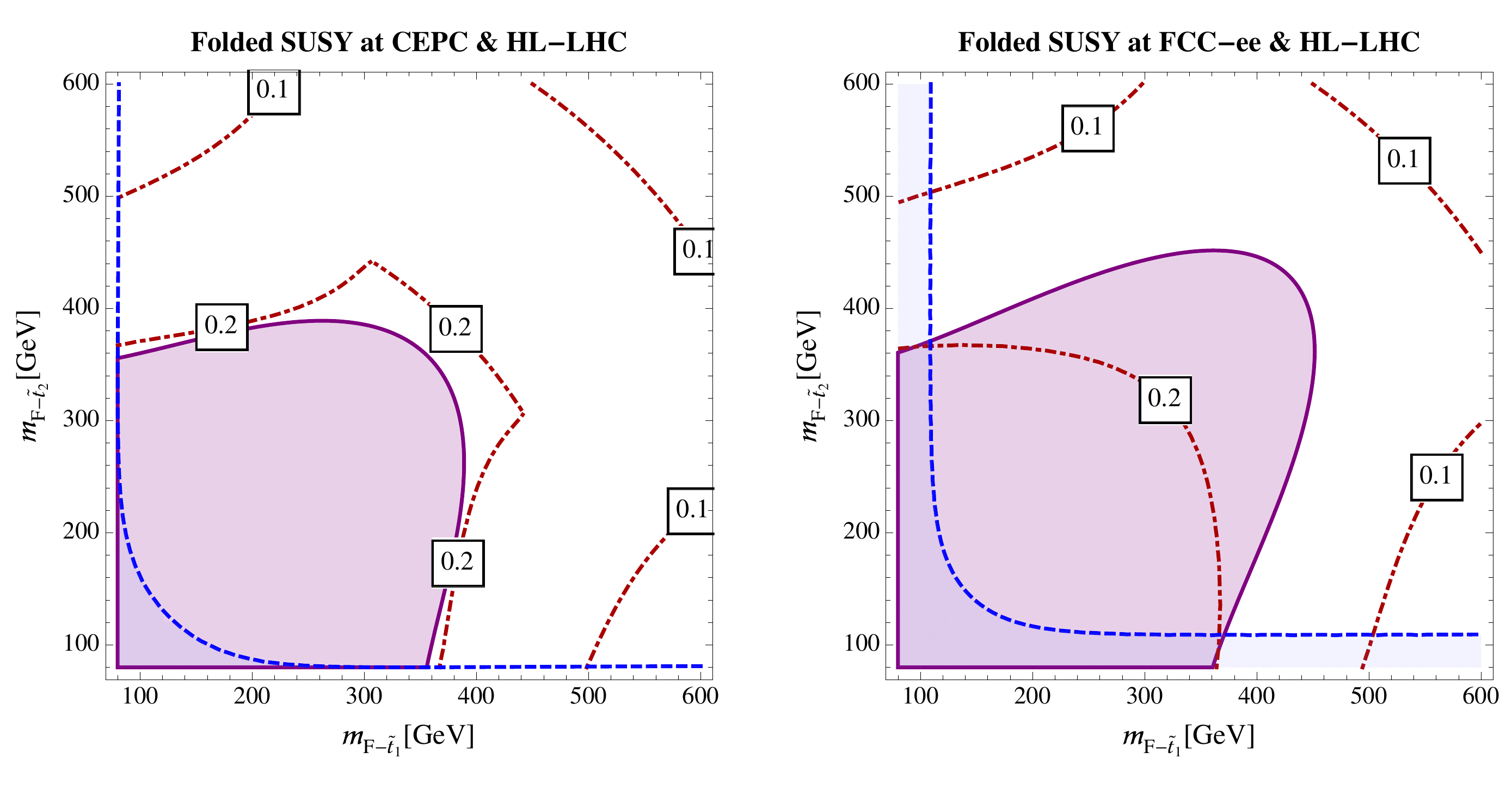}
\end{center}
\caption{Projected constraints in the {\em folded} stop mass plane from a one-parameter fit to the Higgs--photon--photon couplings from future experiments. Directly analogous to Fig.~\ref{fig:stopmassexclude}. Results from the ILC 250/500/1000 would be similar to CEPC; lower-energy ILC measurements provide even weaker constraints. These constraints are subdominant to the constraints on left-handed folded stops arising from $T$-parameter measurements, which are the same as those for ordinary stops in the left-hand column of Fig.~\ref{fig:result}.}
\label{fig:foldedstopmassexclude}
\end{figure}%
%%%%%%%%%%%%%%%%%%%%%%%%%%%%%%%%%%%%%%%%%%%%%%%%%%%%%

EWPT could be the most sensitive experimental probe in some hidden natural SUSY scenarios such as ``folded SUSY"~\cite{Burdman:2006tz}. In folded SUSY, the folded stops only carry electroweak charges and some beyond SM color charge but no QCD charge. The most promising direct collider signal is $W +$ photons which dominates for the ``squirkonium" (the bound state of the folded squarks) near the ground state~\cite{Burdman:2008ek,Burdman:2014zta}. It is a very challenging experimental signature. Among the Higgs coupling measurements, folded stops could only modify the Higgs--photon coupling, the Higgs--photon--$Z$ coupling, and (at a subleading level) the Higgs--$Z$--$Z$ coupling. Yet the Higgs--photon coupling measurements, even at future $e^+e^-$ colliders, have very limited sensitivities. Even FCC-ee Higgs measurements could only probe folded stops up to 400 GeV, as illustrated in Fig.~\ref{fig:foldedstopmassexclude} (which updates the result in~\cite{Fan:2014txa} to include CEPC). Notice that we have also taken into account of a precise determination of $\Gamma(h\to\gamma\gamma)/\Gamma(h\to ZZ)$ at HL-LHC. It has been demonstrated that combing this with Higgs measurements at future $e^+e^-$ colliders could result in a significant improvement of sensitivity to Higgs--photon--photon coupling~\cite{Han:2013kya, Peskin:2013xra}. 

On the other hand, the reach of the electroweak precision we derived in this article (the left column of Fig.~\ref{fig:result}) applies to folded stops as well as the usual stops. Except for the blind spot in the parameter space, future EWPT could probe left-handed folded stops, via their correction to the $T$ parameter, up to 600 GeV (e.g. at the ILC) or even 1 TeV (e.g. at FCC-ee). CEPC's preliminary plans fall close to the ILC reach, but conceivable upgrades could achieve similar reach to FCC-ee. These EWPT constraints would surpass the Higgsstrahlung constraints on folded SUSY estimated in ref.~\cite{Craig:2014una}. Improved measurements of the $W$ mass, then, may be one of the most promising routes to obtaining stronger experimental constraints on folded SUSY. Therefore, with the help of future electroweak  precision measurements, we can test the fine tuning of folded SUSY at the few percent level.

%%%%%%%%%%%%%%%%%%%%%%%%%%%%%%%%%%
\section*{Acknowledgments}
%%%%%%%%%%%%%%%%%%%%%%%%%%%%%%%%%%
We thank Maxim Perelstein and Witek Skiba for useful discussions and comments. We thank the CFHEP in Beijing for its hospitality while this work was initiated and a portion of the paper was completed. The work of MR is supported in part by the NSF Grant PHY-1415548. L-TW is supported by the DOE Early Career Award under Grant DE-SC0003930.
\appendix
%%%%%%%%%%%%%%%%%%%%%%%%%%%%%%%%%%
\section{Formulas for loop effects}
\label{app:loops}
%%%%%%%%%%%%%%%%%%%%%%%%%%%%%%%%%%

\subsection{$S$ parameter}

The $S$-parameter contribution of stops and sbottoms from ref.~\cite{Drees:1991zk} can be simplified to:
\beq
S & = & \frac{1}{8\pi} \left[\left(6 \cos^4 \thetast  - 8 \cos^2 \thetast\right) b(\mst[1], \mst[1]) +  \left(6 \sin^4 \thetast - 8 \sin^2 \thetast\right) b(\mst[2], \mst[2]) + 3 \sin^2(2\thetast) b(\mst[1],\mst[2]) \right. \nonumber \\ & & \left. +\left(6 \cos^4 \thetasb - 4 \cos^2\thetasb\right) b(\msb[1], \msb[1]) + \left( 6 \sin^4 \thetasb - 4 \sin^2 \thetasb\right) b(\msb[2], \msb[2]) + 3\sin^2(2\thetasb) b(\msb[1], \msb[2]) \right]. \nonumber \\
& &
\eeq
where 
\beq
b(x,y) & \equiv & \frac{2}{3(x^2 - y^2)^3} \left(x^4 (x^2 - 3 y^2) \log x + y^4 (3 x^2 - y^2) \log y\right) - \frac{5 x^4 - 22 x^2 y^2 + 5 y^4}{18\left(x^2 - y^2\right)^2} \nonumber \\
& = & \frac{2}{3} \log x~~{\rm if}~~x = y.
\eeq
In particular, in the case of unmixed left-handed stops and sbottoms (split only by Yukawas and $D$-terms), we have
\beq
S = \frac{1}{6\pi} \log \frac{m_{{\tilde b}_L}}{m_{{\tilde t}_L}}  \approx -\frac{1}{6 \pi} \frac{m_t^2}{m_{{\tilde Q}_3}^2}.
\eeq

\subsection{$T$ parameter}

For simplicity, we decouple right-handed sbottoms and assume a negligible sbottom mixing. Then the $T$ parameter is~\cite{Drees:1991zk}: 
\beq
T_{\tilde{t}} =\frac{3\cos^2{\theta_{\tilde{t}}}}{16\pi s_W^2 m_W^2}\left(-\sin^2{\theta_{\tilde{t}}}F_0(m_{\tilde{t}_1}^2,m_{\tilde{t}_2}^2)+F_0(m_{\tilde{t}_1}^2,m_{\tilde{b}_1}^2)+\tan^2\theta_{\tilde{t}}F_0(m_{\tilde{t}_2}^2,m_{\tilde{b}_1}^2)\right),
\label{eq:Tstop}
\eeq
where $m_{\tilde{t}_1}^2, m_{\tilde{t}_2}^2, m_{\tilde{b}_1}^2$ are the two physical stop masses squared and the left-handed sbottom mass squared correspondingly. The function $F_0$ is defined as
\beq
F_0(x,y) = x+y - \frac{2x y}{x-y} \log \frac{x}{y}. 
\eeq

\subsection{$R_b$}  
\label{app:Rb}

The full formula for the stop-Higgsino contribution (a vertex correction to the $Z b_L {\overline b}_L$ coupling) could be found in~\cite{Boulware:1991vp, Wells:1994cu}. It involves several Passarino-Veltman integrals and does not simplify as nicely as the results for $S$ and $T$. As a result, we will only present the expansion of the formula in the limit of small higgsino masses relative to the stop masses. These simple analytic formulas can be compared with our discussion in Sec.~\ref{subsec:Rbops} to explain the structure of the most important terms from an effective field theory (EFT) perspective. Expanding the full formula in the limit of small higgsino masses, we find:
\beq
\Delta R_b^{\rm SUSY} & \approx R_b^{\rm SM}(1-R_b^{\rm SM})& \frac{\alpha}{2\pi \sin^2\theta_W}\frac{v_L\lambda_L^2}{v_L^2+v_R^2} \left[\sin \theta_{\tilde{t}}^2 \frac{m_Z^2}{m_{\tilde{t}_1}^2}\left(\frac{4}{9}-\frac{25}{27}\sin^2\theta_W+\frac{1-2\sin^2\theta_W}{3}\log\frac{\mu^2}{m_{\tilde{t}_1}^2}\right)\right.  \label{eq:deltaRbformula}\nonumber \\
&&+ \left.\cos^2 \theta_{\tilde{t}} \frac{m_Z^2}{m_{\tilde{t}_2}^2}\left(\frac{4}{9}-\frac{25}{27}\sin^2\theta_W+\frac{1-2\sin^2\theta_W}{3}\log\frac{\mu^2}{m_{\tilde{t}_2}^2}\right)\right. \nonumber \\
&&+ \left.\cos^2 \theta_{\tilde{t}} \sin^2 \theta_{\tilde{t}}\left(-\frac{1}{2}+\frac{m_{\tilde{t}_1}^2+m_{\tilde{t}_2}^2}{2(m_{\tilde{t}_1}^2-m_{\tilde{t}_2}^2)}\log\frac{m_{\tilde{t}_1}}{m_{\tilde{t}_2}}\right)+ \cdots \right ],
\eeq
\beq
v_L=-\frac{1}{2}+\frac{1}{3}\sin^2\theta_W \quad v_R=\frac{1}{3}\sin^2\theta_W \quad \lambda_L = \frac{m_t}{\sqrt{2}m_W \sin \beta},
\eeq
where $R_b^{\rm SM}\approx0.22$ and we only kept the leading terms in expansions of $\mu^2/m_{\tilde{t}}^2$ and $m_Z^2/m_{\tilde{t}}^2$. We also neglected terms proportional to the bottom Yukawa coupling. The largest terms in $\Delta R_b^{\rm SUSY}$ are the logarithmic terms such as $\cos^2 \theta_{\tilde{t}} \sin^2 \theta_{\tilde{t}} \frac{m_{\tilde{t}_1}^2+m_{\tilde{t}_2}^2}{2(m_{\tilde{t}_1}^2-m_{\tilde{t}_2}^2)}\log\frac{m_{\tilde{t}_1}}{m_{\tilde{t}_2}}$ and $y_t^2\cos^2 \theta_{\tilde{t}}\frac{m_Z^2}{m_{\tilde{t}_2}^2}\log\frac{\mu^2}{m_{\tilde{t}_R}^2}$. The logarithms here suggest that there should be an explanation of these terms based on RG-induced mixing of different dimension-6 operators.

The term proportional to $(m_Z^2/m_{{\tilde t}_2}^2) \log(m_{{\tilde t}_2}^2/\mu^2)$ is precisely the one that we found by first integrating out right-handed stops to generate the operator $\left({\tilde H}_u \cdot Q_3\right)\left(Q_3^\dagger \cdot {\tilde H}_u^\dagger\right)$, then integrating out higgsinos, in eq.~\ref{eq:Rbnomixing} of Sec.~\ref{subsec:Rbops}. The log can be understood as an operator mixing effect when running between the scale where the stops are integrated out and the scale where the higgsinos are integrated out. Furthermore, we have
\beq
\cos^2 \theta_{\tilde t} \sin^2 \theta_{\tilde t}  \propto \sin^2(2 \theta_{\tilde t}) = \left(\frac{2 m_t X_t}{m_{{\tilde t}_2}^2 - m_{{\tilde t}_1}^2}\right)^2,
\eeq
so the term proportional to $\cos^2 \theta_{\tilde t} \sin^2 \theta_{\tilde t} \log(m_{{\tilde t}_1}/m_{{\tilde t}_2})$ is precisely the term that we found from the EFT viewpoint in eq.~\ref{eq:Rbmixing} by first integrating out left-handed stops to generate the operator $\left(h^\dagger i \overleftrightarrow{D}_\mu h\right)\left({\tilde t}_R^\dagger i \overleftrightarrow{D}^\mu {\tilde t}_R\right)$ and then integrating out right-handed stops and higgsinos. The leading terms in the full vertex diagram calculation could all be derived from EFT arguments by integrating out left-handed stops, right-handed stops, and higgsinos in the correct order.

\bibliography{ref}
\bibliographystyle{utphys}
\end{document}